\documentclass[apj]{emulateapj}

\begin{document}
\title{
	Observation of Diffuse Cosmic and Atmospheric Gamma Rays \\
	at Balloon Altitudes with an Electron-Tracking Compton Camera
}

\author{
	Atsushi Takada\altaffilmark{1, 2}, 
	Hidetoshi Kubo\altaffilmark{2},
	Hironobu Nishimura\altaffilmark{2}, 
	Kazuki Ueno\altaffilmark{2}, 
	Kaori Hattori\altaffilmark{2}, 
	Shigeto Kabuki\altaffilmark{2}, 
	Shunsuke Kurosawa\altaffilmark{2}, 
	Kentaro Miuchi\altaffilmark{2}, 
	Eiichi Mizuta\altaffilmark{3}, 
	Tsutomu Nagayoshi\altaffilmark{2}, 
	Naoki Nonaka\altaffilmark{1}, 
	Yoko Okada\altaffilmark{2}, 
	Reiko Orito\altaffilmark{2}, 
	Hiroyuki Sekiya\altaffilmark{2}, 
	Atsushi Takeda\altaffilmark{2}, 
	and Toru Tanimori\altaffilmark{2}
}

\email{kubo@cr.scphys.kyoto-u.ac.jp}

\altaffiltext{1}{Research and Operation Office for Scientific Ballooning, ISAS, JAXA, Yoshinodai 3-1-1, Chuo-ku, Sagamihara, Kanagawa 252-5210, Japan}
\altaffiltext{2}{Department of Physics, Graduate School of Science, Kyoto University, Kitashirakawa-Oiwakecho, Sakyo-ku, Kyoto 606-8502, Japan}
\altaffiltext{3}{Electronic Components and Devices Group, ISAS, JAXA,Yoshinodai 3-1-1, Chuo-ku, Sagamihara, Kanagawa 252-5210, Japan}

\begin{abstract}
We observed diffuse cosmic and atmospheric gamma rays at balloon altitudes with the Sub-MeV gamma-ray Imaging Loaded-on-balloon Experiment I
 (SMILE-I) as the first step toward a future all-sky survey with a high sensitivity.
SMILE-I employed an electron-tracking Compton camera comprised of a gaseous electron tracker as a Compton-scattering target and a scintillation camera as an absorber.
The balloon carrying the SMILE-I detector was launched from the Sanriku Balloon Center of 
the Institute of Space and Astronautical Science/Japan Aerospace Exploration Agency
(ISAS/JAXA) on September 1, 2006, 
and the flight lasted for $6.8$ hr, including level flight for $4.1$ hr at an altitude of $32$--$35$ km. 
During the level flight, 
we successfully detected $420$ downward gamma rays between $100$ keV and $1$ MeV at zenith angles below $60$ degrees.
To obtain the flux of diffuse cosmic gamma rays,
we first simulated their scattering in the atmosphere using Geant4,
and for gamma rays detected at an atmospheric depth of $7.0$ g cm$^{-2}$, 
we found that $50$\% and $21$\% of the gamma rays at energies of $150$ keV and $1$ MeV, respectively, 
were scattered in the atmosphere prior to reaching the detector.
Moreover, by using Geant4 simulations and the QinetiQ atmospheric radiation model, 
we estimated that the detected events consisted of diffuse cosmic and atmospheric gamma rays ($79$\%),
secondary photons produced in the instrument through the interaction between cosmic rays and materials surrounding the detector ($19$\%),
and other particles ($2$\%).
The obtained growth curve was comparable to Ling's model,
and the fluxes of diffuse cosmic and atmospheric gamma rays were consistent with the results of previous experiments.
The expected detection sensitivity of a future SMILE experiment
 measuring gamma rays between $150$ keV and $20$ MeV was estimated from our SMILE-I results
and was found to be ten times better than that of other experiments at around $1$ MeV.
\end{abstract}

\keywords{balloons --- instrumentation: detectors --- diffuse radiation --- gamma rays: observations}

\section{Introduction}
Observations in the low-energy gamma-ray band from hundreds of keV to tens of MeV 
provide a unique window on such phenomena as 
gamma-ray lines from nuclear de-excitation produced by nucleosynthesis in supernovae, e.g., $^{56}$Ni, $^{56}$Co, and $^{44}$Ti
\citep{Matz_1988, Chevalier_1992, Ballmoos_1995, Schonfelder_2001, Cheng_2004, Boggs_2006}; 
gamma-ray lines from long-lived isotopes spread throughout our galaxy, such as $^{26}$Al and $^{60}$Fe \citep{Oberlack_1996, Knodlseder_1999, Diehl_2003};
electron--positron annihilation lines from the Galactic center \citep{Purcell_1997, Knodlseder_2003, Weidenspointner_2006}; 
neutron capture lines from solar flares \citep{Share_2000};
and gamma-ray resonant absorption along lines of sight toward gamma-ray bright quasars due to delta-resonance or giant dipole resonance \citep{Iyudin_2005}.
Furthermore, 
radiation from black hole accretion disks has been detected \citep{McConnell_2002},
and radiation of neutral pions produced 
by ions accelerated in the strong gravitational potential of black holes 
is also expected to be detected \citep{McConnell_1996, Mahadevan_1997, Battacharyya_2003}.
We can also observe nonthermal processes in low-energy gamma-rays: 
synchrotron radiation or inverse Compton-scattered gamma rays 
in gamma-ray pulsars \citep{Aharonian_1998, Thompson_1999, Schonfelder_2000, Kuiper_2001}, 
active galactic nuclei \citep[AGN;][]{Urry_1995, Fossati_1998, Ghisellini_1998, Kubo_1998, Chiaberge_2001}, 
and gamma ray bursts \citep[GRB;][]{Briggs_1999, Paciesas_1999, Kaneko_2006}.
In addition, 
there are diffuse galactic gamma rays probably produced by bremsstrahlung and inverse Compton scattering of electrons \citep{Boggs_2000, Strong_2000}, 
and diffuse extragalactic gamma rays, which are thought to be a combination of emissions from AGNs and Type Ia supernovae \citep{The_1993, Watanabe_1999} 
or the combined emission from the Comptonization process
including nonthermal electrons in accretion disk coronae of AGNs \citep{Inoue_2008}.
In addition, the measurement of photon polarization through low-energy gamma-ray observations from nonthermal processes,
such as synchrotron radiation and Compton scattering, 
would present a powerful diagnostic tool for astrophysics \citep{Lei_1997, Forot_2008}.
However, observation of this low-energy gamma-ray band is difficult because of the following reasons: 
the dominant process in a detector is Compton scattering
and large backgrounds of photons are produced 
in the hadronic process between cosmic rays and a satellite body. 
Therefore, MeV gamma-ray astronomy has not advanced in comparison with X-ray or other gamma-ray bands. 
In fact, COMPTEL onboard the Compton Gamma Ray Observatory ({\it CGRO}) 
discovered only $\sim30$ steady gamma-ray sources in the $0.75$--$30$ MeV band \citep{Schonfelder_2000}, 
whereas EGRET detected $\sim270$ sources \citep{Hartman_1999},
and {\it Fermi} found $1451$ sources during the first 11 months of the all-sky survey \citep{Abdo_2010} 
in the sub-GeV/GeV region above $100$ MeV.

COMPTEL, which was the first Compton telescope onboard a satellite, 
localized the direction of an incident gamma ray on an event circle superposed on the sky 
by measuring the direction of a Compton-scattered gamma ray and the energies of both the scattered gamma ray and a Compton-recoil electron.
After the launch,
the sensitivity of COMPTEL was found to be mainly determined not by diffuse cosmic gamma rays
but by 1) locally produced background gamma rays; 
2) non-Compton-scattering events such as multiple-photon events due to hadronic interaction, bremsstrahlung, or annihilation of positrons;
and 3) events due to other particles such as neutrons,
even after effective background rejection by time-of-flight measurement 
between a Compton scattering material and an absorber of the scattered gamma rays \citep{Weidenspointner_2001}.
COMPTEL's limited sensitivity made it desirable to create a new instrument with a higher sensitivity in the sub-MeV/MeV region.
Therefore, we developed an electron-tracking Compton camera (ETCC) 
consisting of a three-dimensional tracker of Compton-recoil electrons and an absorber of Compton-scattered gamma rays 
by a new detection method with powerful background rejection \citep{Orito_2003, Tanimori_2004}. 
Compared with COMPTEL, the ETCC can restrict the direction of an incident gamma ray to a reduced arc on the Compton circle
by measuring the three-dimensional track of the Compton-recoil electron 
in addition to the energy of the Compton-recoil electron and the energy and direction of the Compton-scattered gamma ray.
Moreover, the angle between the direction of the scattered gamma ray and that of the Compton-recoil electron 
can be used for powerful background rejection by checking the consistency between the measured and kinematically calculated values.
The ETCC can also detect the polarization of an incident gamma ray, 
because the azimuth angle distribution of Compton scattering has an asymmetry for polarized gamma rays.

Because we had already confirmed the detection principle of the ETCC by ground-based experiments \citep{Takada_2005, Hattori_2007}, 
we initiated an experiment to observe sub-MeV gamma rays from celestial objects by using a balloon-borne camera,
the Sub-MeV gamma-ray Imaging Loaded-on-balloon Experiment (SMILE).
For the first flight of SMILE (hereafter SMILE-I), 
in order to study the background gamma rays for observations of celestial objects and verify the background rejection capability,
we observed both 
atmospheric gamma rays, generated by the interaction between cosmic-ray
particles and nuclei in the atmosphere, and diffuse cosmic gamma rays 
at an altitude of up to $\sim 35$ km \citep{Takada_2007}. 

Since the 1960s, several experiments have been conducted to observe diffuse cosmic gamma rays. 
The first detection was using the {\it Ranger 3} satellite between $70$ keV and $1.2$ MeV in 1962 \citep{Metzger_1964}.
Thereafter,
the existence of these rays was confirmed using the satellites {\it ERS-18} between $0.25$ and $6$ MeV \citep{Vette_1970},
{\it OSO-3} between $7$ and $100$ keV \citep{Schwartz_1974},
{\it Apollo 15} in the energy range of $0.3$--$27$ MeV \citep{Trombka_1973},
and {\it Kosmos 461} between $28$ keV and $4.1$ MeV \citep{Mazets_1975},
and by using several balloons \citep{Bleeker_1970, Damle_1971, Makino_1975, Kinzer_1978}.
While these gamma-ray observatories used phoswich counters,
observations by using a balloon-borne double-Compton telescope in the energy range above $1$ MeV 
began in the 1970s \citep{Schonfelder_1977, Schonfelder_1980}. 
Subsequently, 
observations by using satellites have continued with, for example,  
the High Energy Astronomy Observatory \citep[{\it HEAO};][]{Marshall_1980, Kinzer_1997},
the Solar Maximum Mission \citep[{\it SMM};][]{Watanabe_1999}, 
{\it CGRO} \citep{Weidenspointner_2000, Kappadath_1996}, 
and {\it INTEGRAL} \citep{Churazov_2007} that is based on coded aperture imaging.
An MeV bump, which is in excess near a few MeV in the {\it Apollo} observations, remained a mystery for a long time but is now rejected,
because COMPTEL revealed 
that it was an artifact from delayed gamma-rays produced in radioactive decays in the instrument \citep{Weidenspointner_2000}.

Atmospheric gamma rays have been observed by balloon experiments since the 1960s \citep{Peterson_1972, Kinzer_1974, Ryan_1977, Schonfelder_1980}.
Low-energy atmospheric gamma rays are mainly produced by bremsstrahlung from secondary cascade electrons \citep{Peterson_1973, Danjo_1972},
and the electron--positron annihilation line is also detected \citep{Peterson_1963, Ling_1977}.
Some models for atmospheric gamma rays exist, for example,
the Ling model \citep{Ling_1975} and the QinetiQ Atmospheric Radiation Model \citep[QARM;][]{Lei_2006}.
Previous observations showed
that the flux of atmospheric gamma rays depends on the zenith angle, 
and models produced by the following scientists are based on the observational results: 
Ling, Costa \citep{Costa_1984}, and Graser \citep{Graser_1977}.

Cosmic and atmospheric charged particles and neutrons could present large backgrounds in gamma-ray observations; 
thus, the estimation of the fluxes of these particles is essential \citep{Wunderer_2006, Bowen_2007}. 
The fluxes of these particles fluctuate on the basis of solar activity, observation altitude, and cutoff rigidity. 
A few models for energy spectra of these particles at balloon altitudes have been proposed, such as Mizuno's model \citep{Mizuno_2004} and QARM.

Recently, in addition to an ETCC,
balloon experiments with new detection methods have been performed 
to achieve higher sensitivity in the sub-MeV/MeV band. 
For example, a double-Compton camera using a liquid Xe detector: the LXeGRIT \citep{Aprile_2008}; 
multiple Compton cameras \citep{Kamae_1987} by using semiconductors: 
the nuclear Compton telescope \citep[NCT;][]{Boggs_2007}, a Si/CdTe Compton telescope \citep{Takahashi_2005}, 
and the Tracking and Imaging Gamma Ray Experiment \citep[TIGRE;][]{Zych_2008}; 
an electron-tracking Compton camera using an electron tracker made of a semiconductor: 
the Medium Energy Gamma-ray Astronomy (MEGA) telescope \citep{Bloser_2006}; 
and a gamma-ray lens: CLAIRE \citep{Ballmoos_2005}. 
Also, such experiments have renewed the attention given to observations of diffuse cosmic gamma rays, atmospheric gamma rays,
and the background formed by cosmic rays at balloon altitudes.

In this paper, we report the observational results of SMILE-I.
We first reconstructed the gamma-ray events and then 
estimated the background produced by the interaction between cosmic ray and the materials surrounding the detector during the flight.
Next, we simulated the scattering of diffuse cosmic gamma rays in the atmosphere.
After that we obtained the fluxes of diffuse cosmic and atmospheric gamma rays measured with SMILE-I.
Finally, we determined the detection sensitivity of SMILE-I,
and estimated the expected detection sensitivity of a future SMILE mission.

\section{Instrument}
\begin{figure}
	\plotone{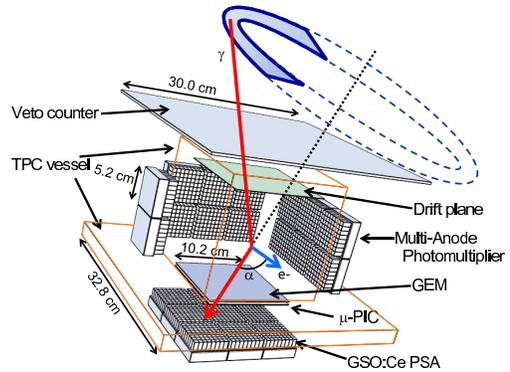}
	\caption{
		Schematic view of our electron-tracking Compton camera (ETCC).
		The ETCC consists of a gaseous tracker that detects the Compton-recoil electrons 
		and a GSO:Ce pixel scintillator array (PSA) that detects the Compton-scattered gamma rays.
		The distance between the drift plane and the gas electron multiplier (GEM) is $14$ cm,
		and that between the GEM and the micro-pixel chamber ($\mu$-PIC) is $0.5$ cm.
		The effective area of the time projection chamber (TPC) itself is the area between the drift plane and GEM.
		\label{fig:mev_camera}
	}
\end{figure}

Figure~\ref{fig:mev_camera} shows a schematic of the ETCC developed by us.
The electron tracker detects the three-dimensional track and the energy of the Compton-recoil electron,
while the absorber detects the absorption point and energy of the Compton-scattered gamma ray. 
By summing the momenta of both the recoil electron and the scattered gamma ray,
we obtain the momentum of the incident gamma ray as
\begin{equation}
	p_0^\mu = p_\gamma^\mu + p_e^\mu ,
\end{equation}
where $p_0^\mu$, $p_\gamma^\mu$, and $p_e^\mu$ are the four-dimensional momenta of the incident gamma ray,
scattered gamma ray, and recoil electron, respectively.
Thus, we can obtain a completely ray-traced gamma-ray image.
The unit vector of the incident gamma ray ${\bf r}$ is described by
\begin{eqnarray}
	{\bf r}
		&= \left( \cos\phi - \frac{\sin\phi}{\tan \alpha} \right) {\bf g} 
			+ \frac{\sin\phi}{\sin\alpha} {\bf e} \\
		&= \frac{E_\gamma}{E_\gamma + K_e} {\bf g} 
			+ \frac{\sqrt{K_e (K_e + 2 m_e c^2)}}{E_\gamma + K_e} {\bf e} \label{eq:kin_rcs},
\end{eqnarray}
where $\bf g$ and $\bf e$ are unit vectors in the directions of the scattered gamma ray and the recoil electron, respectively,
and $\phi$ is the scattering angle given by
\begin{equation}
	\label{eq:phi}
	\cos \phi = 1 - m_e c^2 \left( \frac{1}{E_\gamma} - \frac{1}{E_\gamma + K_e} \right) ,
\end{equation}
where $E_\gamma$, $K_e$, $m_e$, and $c$ are the energy of the scattered gamma ray, kinetic energy of the Compton-recoil electron, 
electron mass, and light speed, respectively.
$\alpha$, which is the angle between the scattering direction and the recoil direction, as shown in Fig.~\ref{fig:mev_camera},
is geometrically measured for each incident gamma ray,
\begin{equation}
	\cos \alpha_{geo} = {\bf g} \cdot {\bf e} ,
\end{equation}
and this angle is also obtained from a calculation by the Compton scattering kinematics:
\begin{equation}
	\cos \alpha_{kin} = \left( 1 - \frac{m_e c^2}{E_\gamma} \right)
		\sqrt{\frac{K_e}{K_e + 2 m_e c^2}} .
\end{equation}
By comparing $\alpha_{geo}$ with $\alpha_{kin}$,
\begin{equation}
	\label{eq:alpha_cut}
	\left| 1 - \frac{\cos \alpha_{geo}}{\cos \alpha_{kin}} \right| \leq \Delta_\alpha ,
\end{equation}
where $\Delta_\alpha$ is a cut parameter,
we can select only Compton scattering events. 
Thus, Compton imaging with electron tracking is an effective method for MeV gamma-ray astronomy, 
with the serious problem of a large background, as described in Section~1.

For SMILE-I,
we have constructed an ETCC with a detection energy range $100$ keV--$1$ MeV.
Our ETCC consists of a gaseous tracker for detecting Compton-recoil electrons by using gas avalanche detectors to read it.
In addition, it has a pixel scintillator array (PSA) for detecting Compton-scattered gamma rays.
We provide a detailed description of our ETCC design in the following paragraphs.
As a tracker for detecting Compton-recoil electrons, we developed a gaseous time projection chamber \citep[TPC;][]{Kubo_2003, Miuchi_2003}
with a volume of $10 \times 10 \times 14$ cm$^3$ filled 
with a gas mixture of $80$\% Xe, $18$\% Ar, and $2$\% C$_2$H$_6$ in mass ratio and sealed at $1$ atm.
The readout of the TPC consists of gas avalanche detectors: a gas electron multiplier \citep[GEM;][]{Sauli_1997, Tamagawa_2006} 
and a micro-pixel chamber \citep[$\mu$-PIC;][]{Ochi_2001, Nagayoshi_D}. 
The latter is our original gaseous two-dimensional imaging detector 
with micro-pixel electrodes produced using printed circuit board technology.
A seed electron drifts into the $50$-$\mu$m-thick GEM with a velocity of $2.5$ cm $\mu$s$^{-1}$ at an electric field of $400$ V cm$^{-1}$,
and the first multiplication is caused by the GEM with a gain of approximately $10$. 
Then, the multiplied electrons drift to the $\mu$-PIC, 
and the second multiplication is caused by the $\mu$-PIC with a gain of $3 \times 10^3$. 
Therefore, we obtain a high gas gain above $3 \times 10^4$, which is enough to detect Compton-recoil electrons.
The signals from the $\mu$-PIC are read using amplifier-shaper-discriminator chips \citep{Sasaki_1999}, 
which feed an output signal from a charge amplifier to a flash analog-to-digital converter to measure the recoil-electron energy,
and a discriminated digital signal to a position encoder 
that encodes the electron track by using field-programmable gate arrays with a 100-MHz clock \citep{Kubo_2005}.
Because the $\mu$-PIC consists of fine-structure pixels with a pitch of $400$ $\mu$m,
this TPC has a fine three-dimensional position resolution of $\sqrt{\sigma_{\perp}^2 + \sigma_{\parallel}^2} = 500$ $\mu$m,
where $\sigma_\perp$ and $\sigma_{\parallel}$ are resolutions perpendicular and parallel to the drift direction, respectively \citep{Takada_2007}.
As an absorber for detecting Compton-scattered gamma rays, we used $33$-pixel-scintillator arrays \citep{Nishimura_2007}, 
each of which consists of $8 \times 8$ GSO:Ce scintillator pixels with a pixel size of $6 \times 6 \times 13$ mm$^3$. 
For the photon sensor of the scintillation camera, 
we selected a multi-anode photomultiplier tube (Hamamatsu Photonics, Flat-Panel H8500),
which consists of $8 \times 8$ anode pixels with a pixel size of $6 \times 6$ mm$^2$. 
With chained resistors connecting the anodes to reduce the number of readout channels, 
we obtained the position of a hit pixel by the charge-division method \citep{Sekiya_2006}. 
A plastic scintillator with a size of $30 \times 30 \times 0.3$ cm$^3$ was placed $21$ cm above the GEM 
as a veto counter for reducing triggers by charged particles.
When an energy deposit in the GSO scintillators was over $30$ keV,
a trigger was generated and the data-acquisition system waited for $8$ $\mu$s 
(the maximum time for seed electrons to drift to the $\mu$-PIC) for the TPC signals.
When a trigger from the veto counter was generated, 
the system waited for $100$ $\mu$s (the recovery time of the undershoot of the large signal caused by charged particles) after the veto trigger. 
\begin{figure}
	\plotone{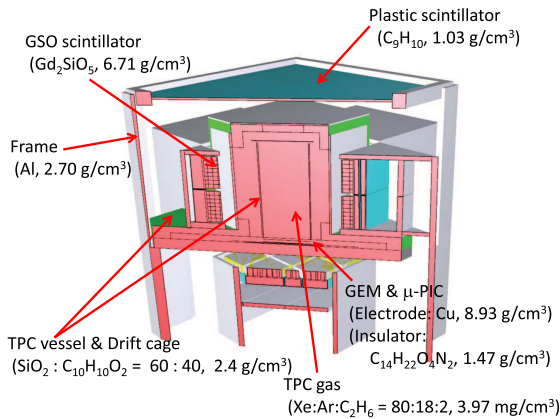}
	\caption{
		Cross-sectional view of a mass model of the SMILE-I ETCC.
		\label{fig:mass_model_etcc}
	}
\end{figure}

\begin{figure}
	\plotone{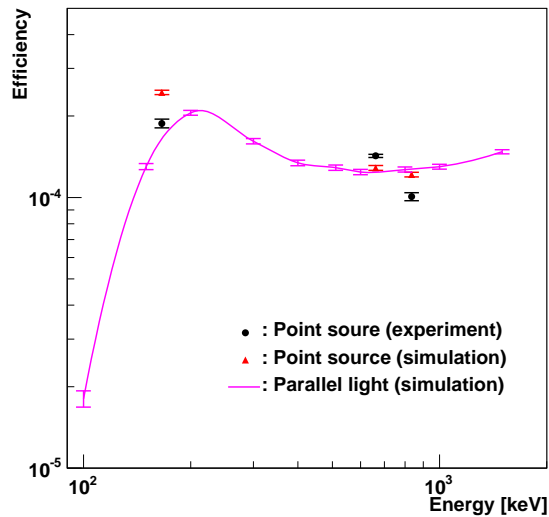}
	\caption{
		Dependence of detection efficiency on the energy of incident gamma rays
		for a point source near the detector, obtained by experiments (circles) and simulation (triangles).
		The dependence of efficiency on the incident energy for parallel light at a zero degree zenith angle
		from the Geant4 simulation using a mass model of Fig.~\ref{fig:mass_model_etcc},
		 is also shown (solid line).
		The error bars represent only statistical uncertainties.
		\label{fig:efficiency}
	}
\end{figure}
\begin{figure}
	\plotone{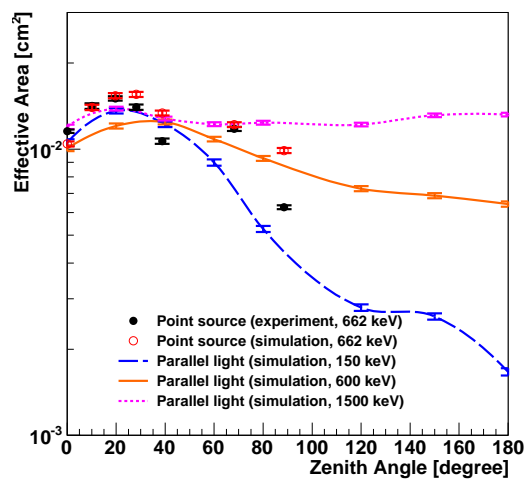}
	\caption{
		Effective area as a function of zenith angle
		for a point source near the detector obtained by experiments (filled circles) and simulations (open circles), 
		and for parallel light with energies of 150 keV (dashed line), 600 keV (solid line), and 1500 keV (dotted line), 
		obtained by Geant4 simulation using a mass model shown in Fig.~\ref{fig:mass_model_etcc}.
		The error bars represent only statistical uncertainties.
		\label{fig:effective_area}
	}
\end{figure}
To obtain the fluxes of diffuse cosmic gamma rays and atmospheric gamma rays by SMILE-I,
the detection efficiency and the effective area for parallel light is required. 
However, irradiating the entire system of SMILE-I by using parallel light in a laboratory is very difficult. 
Instead, we simulated the irradiation with Geant4 \citep[ver 9.0-patch01;][]{Agostinelli_2003} by using a mass model shown in Fig.~\ref{fig:mass_model_etcc}.
In the ETCC simulator, 
the absorber had an energy resolution of $12 \times \left( {\rm Energy} / 662 {\rm keV} \right)^{-0.57}$ \% 
at full width at half maximum (FWHM) and an energy threshold of $30$ keV, 
the tracker had an energy resolution of $45 \times \left( {\rm Energy} / 32 {\rm keV} \right)^{-0.22}$ \% 
at FWHM and an energy threshold of $\sim 3$ keV, 
and the veto counter had an energy resolution of $50$\% at FWHM and an energy threshold of $300$ keV.
The time resolutions of the absorber and the veto counter were not considered, 
because our ETCC did not use the event selection based on the time-of-flight between the Compton scattering target and the absorber.
We obtained the detection efficiency of gamma rays reconstructed 
within zenith angles below $60$ degrees for irradiation from the zenith direction 
as a function of the incident energy, as shown in Fig.~\ref{fig:efficiency}; 
the detection efficiency was approximately $10^{-4}$ between $150$ keV and $1.5$ MeV.
To estimate the systematic uncertainty of the detection efficiency,
we measured and also simulated the detection efficiency for a radioisotope 
under the condition of gamma-ray irradiation from a point source placed at a distance of $50$ cm from the drift plane of the TPC.
The difference between the experimental measurement and simulated data was approximately $15$\%, 
as shown in Fig.~\ref{fig:efficiency}, 
and we incorporated this difference into the systematic uncertainty for the detection efficiency.
In order to verify our mass model simulation of parallel light and to determine if the detection efficiency uncertainty depends on the zenith angle,
we plotted the effective area obtained by simulation for parallel light 
between $150$ keV and $1.5$ MeV as a function of zenith angle (Fig.~\ref{fig:effective_area}).
For the verification of the simulator, we measured and also simulated the effective area for a point radioisotope.
The difference between the simulation and experimental data does not show any systematic dependence on incident direction.
Thus, we verified our mass model simulation, and assumed that the systematic uncertainty of the detection efficiency is not dependent on the zenith angle.
Figure~\ref{fig:effective_area} also shows 
that the acceptance field-of-view (FOV) of SMILE-I was $120$ degrees ($3$ sr) full width at half maximum (FWHM) for $150$ keV,
$200$ degrees ($7$ sr) for $600$ keV, and more than $7$ sr for $1500$ keV, respectively,
which was wider than that of COMPTEL that was $2$ sr (FWHM) at $1.2$ MeV and $0.5$ sr at $6.1$ MeV \citep{Schonfelder_1993}.
Table~\ref{tab:res} shows the energy and the angular resolutions of the ETCC as a function of the incident energy,
measured at a ground-based experiment.
The angular resolution of the ETCC is defined by two parameters: 
the angular resolution measure (ARM), which is the accuracy of the scattering angle $\phi$, 
and the scatter plane deviation (SPD), which is the determination accuracy of the Compton-scattering plane \citep{Bloser_2002}.
In general,
the theoretical limit of the ARM is Doppler broadening caused by the undetectable momentum of the Compton-target electron, 
and the SPD is limited by the multiple scattering of the Compton-recoil electron;
thus, an ETCC using a gas with a higher atomic number has a poorer angular resolution due to the larger multiple scattering,
although it has a higher detection efficiency. 
The ARM resolution due to Doppler broadening is $3.3$ degrees for $200$ keV and $0.7$ degrees for $1$ MeV at FWHM. 
The SPD resolution due to multiple scattering of the recoil electron with recoil energy of tens of keV is approximately $100$ degrees at FWHM.
In fact, an ETCC using an Ar gas had the ARM and SPD of $8.4$ degrees and $89$ degrees (FWHM), respectively, at $662$ keV \citep{Takada_2007},
and these values were better than those obtained using the higher atomic number Xe gas in the SMILE-I ETCC detector.
Because we intended to observe not a celestial point source
but diffuse cosmic gamma rays and atmospheric gamma rays in SMILE-I,
we gave preference to the higher efficiency over the better angular resolution and, thus, adopted Xe gas for our ETCC.
\begin{deluxetable}{cccc}
	\tablecolumns{4}
	\tablewidth{0pc}
	\tablecaption{Energy resolution and angular resolutions (given by the FWHM) of the ETCC \label{tab:res}}
	\tablehead{
		\colhead{Energy} &\colhead{Energy resolution} &\colhead{ARM} &\colhead{SPD} \\
		\colhead{[keV]} &\colhead{[\%]} &\colhead{[degrees]} &\colhead{[degrees]}
	}
	\startdata
	166 &$37\pm2$ &$46\pm2$ &$189\pm17$ \\
	356	&$20\pm2$ &$24\pm2$ &$181\pm10$ \\
	511	&$15\pm1$ &$21\pm2$ &$180\pm15$ \\
	662 &$14\pm1$ &$18\pm1$ &$183\pm13$ \\
	835	&$15\pm2$ &$21\pm4$ &$185\pm17$ 
	\enddata
\end{deluxetable}

For the balloon flight,
the ETCC was placed in an aluminum vessel with a diameter of $1$ m, a height of $1.4$ m, and a thickness of $3$ mm.
The vessel was maintained at $1$ atm 
and the vessel was fixed to an aluminum gondola with a size of $1.2 \times 1.5 \times 1.6$ m$^3$. 
On two sides of the gondola, 
batteries and ballast boxes were attached, and the gondola was packed with expanded polystyrene. 
In addition,
the SMILE-I gondola had a fine pressure gauge for measurement of atmospheric pressure, 
a global positioning system receiver for measurement of the balloon's altitude and geographic position, 
and two clinometers and two geomagnetic aspectmeters for determining attitude.

\section{Balloon Flight}
\begin{figure}
	\plotone{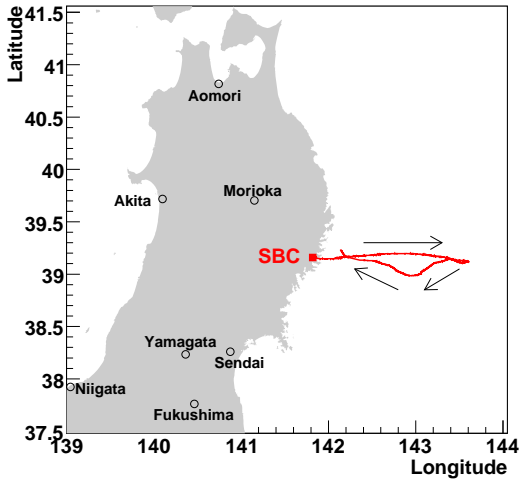}
	\caption{
		Flight path of SMILE-I.
		The balloon was launched from the Sanriku Balloon Center (SBC) of ISAS/JAXA in Japan (39.16N, 141.82E),
		and it landed in the sea approximately $30$ km from the SBC.		
		\label{fig:flight_path}
	}
\end{figure}

\begin{figure}
	\plotone{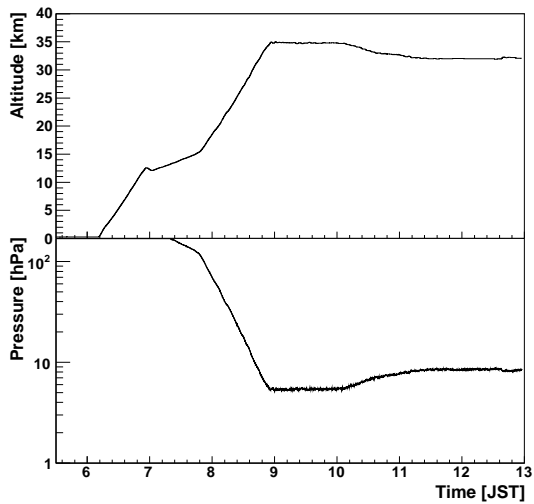}
	\caption{
		Time variation of the balloon's altitude and atmospheric pressure.
		\label{fig:altitude}
	}
\end{figure}

The SMILE-I balloon was launched from 
the Sanriku Balloon Center of the Institute of Space and Astronautical Science/Japan Aerospace Exploration Agency
(ISAS/JAXA; $39.16^\circ$N,  $141.82^\circ$E) on September 1, 2006 at 06:11 Japan Standard Time (JST). 
The flight path is shown in Fig.~\ref{fig:flight_path}.
Figure~\ref{fig:altitude} shows the time variations of the balloon's altitude and atmospheric pressure.
At 08:56, the balloon reached an altitude of $35.0$ km and began level flight. 
From 08:56 to 10:15, 
the altitude was constant at $35$ km and the atmospheric pressure was $5.4$ hPa ($5.5$ g cm$^{-2}$),
and from 11:20 to 13:00, the altitude was $32$ km and the pressure was $8.5$ hPa ($8.7$ g cm$^{-2}$).
During level flight between 08:56 and 13:00, 
the time average of the atmospheric depth was $7.0$ g cm$^{-2}$,
the live time of the observation was $3.0$ hr,
and the cutoff rigidity calculated by QARM was $9.7$ GV.
Between 12:06 and 12:33, 
we suspended gamma-ray observation, and the camera was operated in the charged-particle tracking mode,
which was used to check the performance of the tracker with a trigger of any two hits on GSO pixels.
At 12:33, we resumed gamma-ray observation.
Then, we turned off the system power at 12:59.
Finally, we cut the gondola from the balloon at 13:20. 
The gondola landed in the sea at 13:45, 
and we successfully recovered it at 14:32.
During the whole flight,
no serious problem occurred with the balloon system or the detector. 

\section{Results \& Discussion}
\subsection{Event Reconstruction}
\begin{figure}
	\plotone{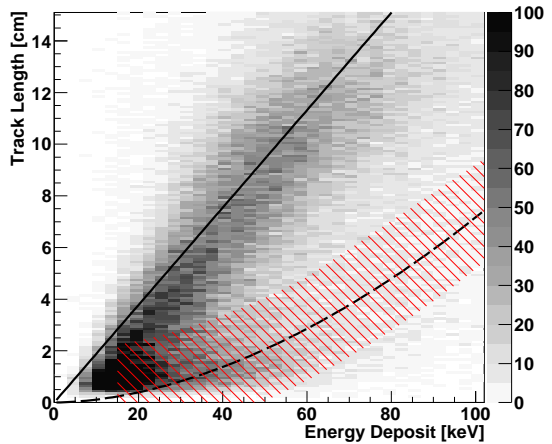}
	\caption{
		Track length and deposited energy in the TPC for events that occurred during level flight (scatter plot).
		The solid and dashed lines represent the relationship calculated from energy deposit per unit length of minimum ionizing particles 
		and the relationship simulated using Geant4 for electrons stopped in the TPC, respectively.
		The hatched area represents the event selection described by equations~(\ref{eq:e_tr}) and (\ref{eq:e_lim}).
		\label{fig:e_tr}
	}
\end{figure}

\begin{figure}
	\plotone{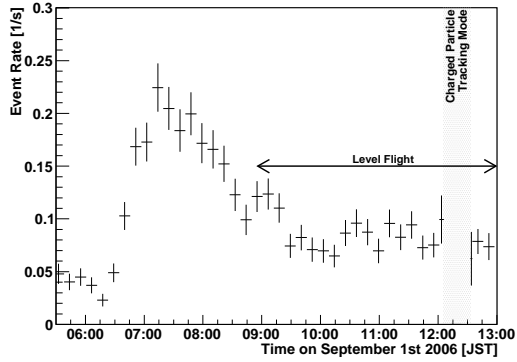}
	\caption{
		Time variation of the rate of reconstruction of events.
		\label{fig:rate_flight}
	}
\end{figure}

\begin{figure}
	\plotone{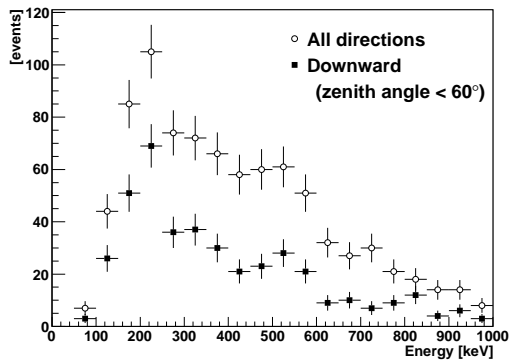}
	\caption{
		Energy spectra of the events that were reconstructed as gamma rays during level flight in all directions (circles),
		and at zenith angles below $60$ degrees (squares). 
		\label{fig:spec_level_flight}
	}
\end{figure}

In this flight, $2.2 \times 10^5$ events were acquired during the $6.8$ hr after the launch.
If an incident gamma ray had been scattered in the TPC and then hit more than two pixels in the GSO:Ce absorber, 
we could not reconstruct the gamma-ray event because the sequence of interactions in the absorber was unknown. 
We, therefore, selected events with a single hit in the absorber,
and $1.1 \times 10^5$ events remained.
Figure~\ref{fig:e_tr} shows the energy deposit in the TPC and the track length of those events.
If a Compton-recoil electron stops in the TPC, 
the event is plotted around the dashed line in this figure, which represents the relationship obtained from the Geant4 simulation.
On the other hand, 
if the electron escaped from the TPC, 
the event is located around the solid line, which represents the relationship calculated from energy deposit per unit length for minimum ionizing particles.
This is because with increasing energy of the recoil electron its energy deposit per unit length of the electron escaping 
from the TPC approaches that of minimum ionizing particles.
Because the energy of a Compton-recoil electron is necessary for the gamma-ray reconstruction, 
$6.5 \times 10^3$ events remained after we selected events in the case that the electron stopped in the TPC.
We used the following criteria to select events in which the electron stopped in the TPC: 
\begin{eqnarray}
	& \left| \frac{L_e}{\rm [cm]} - 1.8\times 10^{-3} \left( \frac{K_e}{\rm [keV]} \right) ^{1.8} \right| \leq 2 \label{eq:e_tr}, \\
	& K_e > 15 \ {\rm [keV]} \label{eq:e_lim},
\end{eqnarray}
where $L_e$ is the length of the detected track in the tracker.
We know from our experience with ground-based experiments using the prototype trackers 
that an electron deposits all its energy in the TPC if it satisfies equation~(\ref{eq:e_tr}).
The energy spectrum of the TPC has a line component at $8$ keV 
which corresponds to the characteristic CuK$\alpha$ X-ray line from the copper electrode of the TPC. 
Because these events are not Compton events, we used the threshold given in equation~(\ref{eq:e_lim}) to reject them.
Thus, we selected Compton events in which the electron stopped in the TPC in order to be able to reconstruct the incident gamma rays.
Finally, we selected events whose Compton scattering occurred in the fiducial volume of $9 \times 9 \times 13$ cm$^3$
by using the angle $\alpha$ and equation~(\ref{eq:alpha_cut}) with $\Delta_\alpha$ of
\begin{equation}
	\Delta_\alpha = 0.075 \times \frac{(E_\gamma + K_e)^2}{K_e \left| E_g - m_ec^2 \right|} \label{eq:d_alpha},
\end{equation}
$2.1 \times 10^3$ events then remained.
This criteria of the angle $\alpha$ is equivalent to $| 1 - |{\bf r}|^2 | \leq 0.15$,
where $\bf r$ was described by equation~(\ref{eq:kin_rcs})
and $|\bf r|$ is equal to unity under the condition of $\cos \alpha_{geo} = \cos \alpha_{kin}$.
The time variation of the count rate of the reconstructed events is shown in Fig.~\ref{fig:rate_flight},
which has a maximum near the Pfotzer maximum
and is thereafter constant during level flight, with a standard deviation of $20$\%.
During level flight, 
we obtained $881$ reconstructed events in the detected energy range of $100$ keV--$1$ MeV, 
$420$ events of which were downward at zenith angle less than $60$ degrees.
The energy spectra of both the downward events and the events in all directions are shown in Fig.~\ref{fig:spec_level_flight}. 

\subsection{Instrumental Background Simulation}
In a balloon-borne experiment, 
particles incident on a detector are not only gamma rays but also protons, electrons, neutrons, and others. 
These particles and the gamma rays induced by interactions between these particles 
and the pressurized vessel or materials surrounding the detector could form a part of the background. 
In particular, 
a neutron causes elastic scattering, 
so that when a neutron interacts with both the scatterer and the absorber of a Compton camera,
the neutron event may be incorrectly reconstructed as a Compton-scattered gamma ray in Compton-type detectors.
The electron tracker of an ETCC can distinguish between electrons stopped in the TPC and other charged particles 
because of the difference in the track length as a function of the kinetic energy, as shown in Fig.~\ref{fig:e_tr}.
The purpose of the SMILE-I experiment was the measurement of diffuse cosmic and atmospheric gamma rays. 
Therefore, atmospheric gamma rays produced in the atmosphere by cosmic-ray interactions are considered not as background but as signals in this experiment.

\begin{figure}
	\plotone{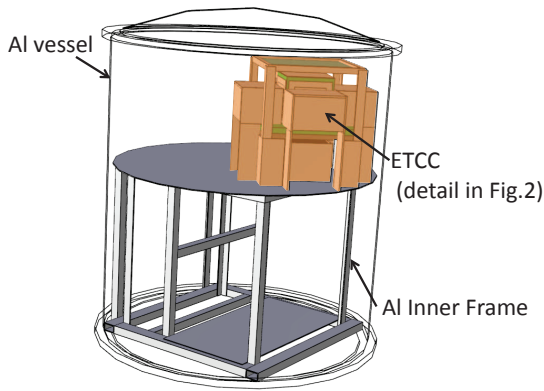}
	\caption{
		Mass model for the instrumental background simulation.
		The model of the electron-tracking Compton camera (ETCC) is the same as in Fig.~\ref{fig:mass_model_etcc}.
		\label{fig:mass_model_vessel}
	}
\end{figure}
To estimate background events, 
we calculated the background radiation at the balloon's altitude by using a Geant4 simulation and a mass model shown in Fig.~\ref{fig:mass_model_vessel}. 
The balloon gondola was not included in the mass model, 
because most of the parts of the balloon gondola were expanded polystyrene and the effect on secondary particle production was negligible.
Due to the low density of the electronics boxes and the large distance between the ETCC and the battery, 
the effect of the instrumental background can be neglected.
Therefore, this mass model does not include a gondola, electronics box, and battery.
The Geant4 hadronic processes were elastic and inelastic scattering for the low- and high-energy models, 
and absorption for pions, kaons, protons, anti-protons, and anti-neutrons; 
elastic and inelastic scattering, and capture with a high-precision model below $19$ MeV for neutrons;
elastic and inelastic scattering with a low-energy model for deutrons, tritons, and alpha particles.
In addition, {\it G4Decay} and {\it G4RadioactiveDecay} processes were included for prompt processes and delayed photons from radioactive decays.
We did not include the chance coincidences between triggers from different components of the radiation background,
since their contribution is expected to be small by the background rejection method described in section 2.

\begin{figure}
	\plotone{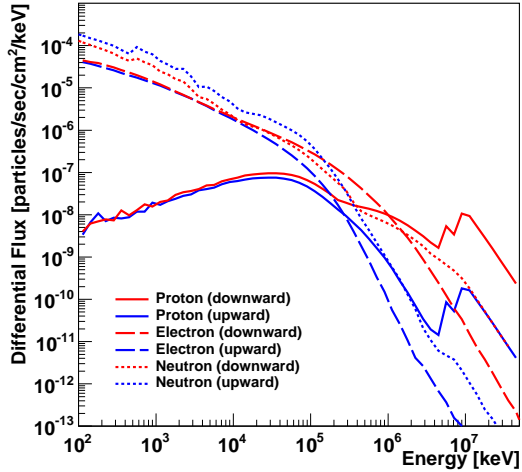}
	\caption{
		Energy spectra of protons (solid lines), electrons (dashed lines), and neutrons (dotted lines) for a background simulation 
		at an atmospheric depth of $7.0$ g cm$^{-2}$.
		Red shows the downward flux in the zenith angle range of 0-90 degrees and blue shows the upward flux in the zenith angle range of 90-180 degrees.
		All fluxes were calculated by QARM.
		\label{fig:back_inc}
	}
\end{figure}

As described in section 1,
there are several radiation models for the fluxes of incident particles at balloon altitudes.
Mizuno's model is a radiation model based on the previous observations 
of protons and alpha particles above $100$ MeV, electrons and positrons above $100$ MeV, 
gamma rays between $1$ MeV and $100$ GeV, and muons between $300$ MeV and $20$ GeV.
However, below these energy ranges, 
the flux of each of these particles is simply assumed to be a power-law spectrum with an index of $-1$,
and the neutron flux is not described at all in this model.
In contrast, the QARM model, based on previous observations and some simulations, 
has fluxes of neutrons in addition to protons, electrons, gamma rays, pions, and muons in wider energy ranges between $100$ keV and $100$ GeV
and includes the effect of solar modulation on the spectral shape and normalization.
The parameters of the QARM model are date, 
geographic position, altitude/residual pressure, magnetic field disturbance index, and the primary-particle spectrum.
Because the model of low-energy charged particles and neutrons was important in sub-MeV gamma-ray observations, 
we adopted the QARM spectra shown in Fig.~\ref{fig:back_inc},
assuming a primary-particle spectrum of galactic cosmic rays \citep{Lei_2004} 
under a magnetic field disturbance index $K_p$ \citep{Mayaud_1980} of $3$
and an atmospheric depth of $7$ g cm$^{-2}$.
The zenith angle distribution of QARM for public use is divided in only two regions: the zenith angle range of $0$--$90$ degrees, and $90$--$180$ degrees.
Figure~\ref{fig:back_inc} represents the downward fluxes in the zenith angle range of $0$--$90$ degrees and the upward fluxes in the zenith angle range of $90$--$180$ degrees.
We estimated an uncertainty in the particle fluxes of QARM to be $20$\% 
by varying the atmospheric depth between $5.3$ g cm$^{-2}$ and $8.5$ g cm$^{-2}$
and the magnetic field disturbance index between $2$ and $4$.

\begin{figure}
	\plotone{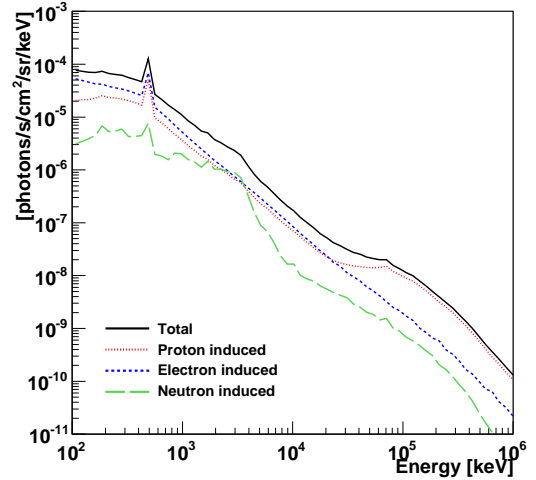}
	\caption{
		Simulated source function of the secondary photons produced in the instrument
		by protons (dotted line), electrons (dashed line), and neutrons (long dashed line)
		at zenith angles between $0$ and $90$ degrees,
		spectra of which are shown in Fig.~\ref{fig:back_inc},
		at an atmospheric depth of $7.0$ g cm$^{-2}$.
		The solid line shows the total flux.
		\label{fig:bg_photon}
	}
\end{figure}

\begin{figure}
	\plotone{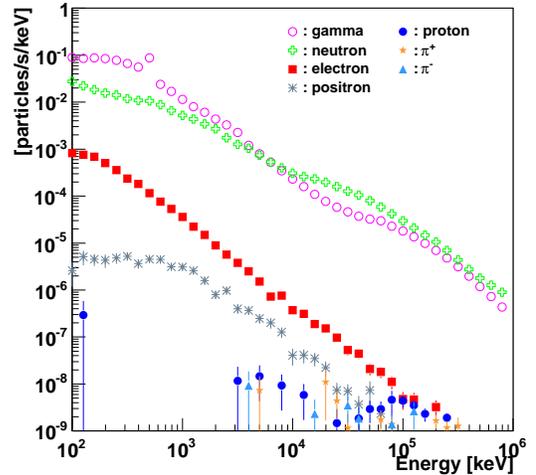}
	\caption{
		Simulated differential fluxes of particles incident on the gaseous tracker from all directions
		for secondary gamma rays (open circles), neutrons (crosses), electrons (squares),
		positrons (asterisks), protons (filled circles), $\pi^+$ (stars), and $\pi^-$ (triangles),
		at an atmospheric depth of $7.0$ g cm$^{-2}$.
		These spectra include both primary (shown in Fig.~\ref{fig:back_inc}) and secondary particles.
		\label{fig:bg_particles}
	}
\end{figure}

\begin{figure}
	\plotone{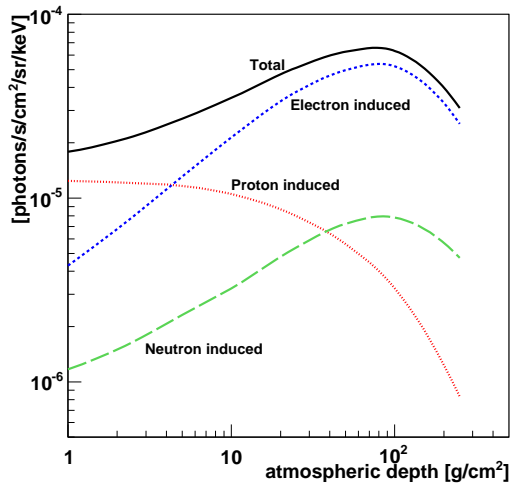}
	\caption{
		Differential flux of instrumental background photons 
		produced by protons (dotted line), electrons (dashed line), and neutrons (long dashed line) in the $125$--$1250$ keV region,
		as a function of atmospheric depth.
		These fluxes represent the background source function averaged in the zenith angle between 0 degree and 90 degrees at the sensitive area of the ETCC.
		The solid line shows the total flux.
		\label{fig:bg_depth}
	}
\end{figure}

For the simulation of the instrumental background coming from any direction,
we assume the incident direction of the primary particles to be downward (zenith angle = $0$ degrees) or upward (zenith angle = $180$ degrees), 
because we do not have any observational data of the horizontal fluxes or the zenith angle distributions of protons, neutrons, and electrons.
For the incident fluxes,
we adopted the QARM flux in the zenith angle range of $0$--$90$ degrees (Fig.~\ref{fig:back_inc}) as the downward primary particle flux,
and the QARM flux in the zenith angle range of $90$--$180$ degrees (Fig.~\ref{fig:back_inc}) as the upward primary particle flux.
The simulated spectra of instrumental background gamma rays produced in the materials surrounding the detector are shown in Fig.~\ref{fig:bg_photon}.
The spectra shown in Fig.~\ref{fig:bg_photon} are the flux averaged over the solid angle in the zenith angle range between $0$ and $90$ degrees.
Gamma rays induced by electrons dominate in the energy range of SMILE-I,
and neutron-induced gamma rays contribute slightly at several MeV. 
In addition, there is a line component of electron--positron annihilation at $511$ keV.
In order to estimate the contribution of secondary gamma rays, neutrons, and charged particles incoming to the gaseous tracker from all directions, 
we simulated their fluxes (Fig.~\ref{fig:bg_particles}).
Throughout this paper, secondary gamma rays refer to photons due to the instrumental background only.
Secondary gamma rays contribute the majority of the background,
with the rest coming from neutrons, electrons, positrons, protons, and charged pions.
The rejection inefficiencies for background particles 
after Compton reconstruction under the constraints of equations~(\ref{eq:e_tr}), (\ref{eq:e_lim}), and (\ref{eq:d_alpha}) were
estimated by the simulation to be $1.8\pm0.3 \times 10^{-5}$ for neutrons, 
$3.4\pm0.2 \times 10^{-4}$ for electrons, and $2.5\pm0.2 \times 10^{-4}$ for protons.
Therefore, 
the simulation predicted that the background event rates of gamma rays, neutrons, and charged particles during level flight would be 
$7.4\pm2.6 \times 10^{-3}$ s$^{-1}$, $6.7\pm2.6 \times 10^{-4}$ s$^{-1}$, and $8.7\pm2.3 \times 10^{-5}$ s$^{-1}$, respectively.
The uncertainties are caused by the uncertainty in the background model.
Because the rate of reconstructed events during the level flight of SMILE-I was $3.9 \times 10^{-2}$ s$^{-1}$, 
the contributions of background gamma rays, neutrons, and charged particles to the obtained events were
$19$\%, $1.7$\%, and $0.23$\%, respectively.
Thus, we estimated that $98.0 \pm 0.7$\% of the reconstructed events were gamma-ray events, 
and the contribution of the other particles was $2.0 \pm 0.7$\%,
thanks to the powerful background rejection of the ETCC.
Figure~\ref{fig:bg_depth} shows the source function of instrumental secondary gamma rays incoming to the detector 
from the zenith direction as a function of atmospheric depth.
These spectra do not include any event selection described in section 4.1.
In later analysis,
we use this flux of gamma rays as the background.

\subsection{Gamma-ray Scattering in the Atmosphere}
In observing diffuse cosmic gamma rays at a few hundred keV, 
there is a contribution of cosmic gamma rays 
that are scattered into the aperture of the detector 
after one or more Compton scatterings in the overlying atmosphere. 
The cosmic gamma-ray flux $f_c (z, \theta)$ at the atmospheric depth of $z$ and the zenith angle of $\theta$ can be separated into two components,
\begin{equation}
	f_c (z, \theta) = f_n (z, \theta) + f_s (z, \theta) .
	\label{eq:cosmic}
\end{equation}
$f_n (z, \theta)$ is the flux of cosmic gamma rays
that enter the aperture of the detector without any interaction in the overlying atmosphere.
In the zenith direction ($\theta = 0$),
the depth dependence of $f_n (z, \theta)$ is given by
\begin{equation}
	f_n (z, 0) = f_n(0, 0) \exp(- \tau_{tot} z) = \eta \exp(- \tau_{tot} z) ,
	\label{eq:cs_att}
\end{equation}
where $\eta$ is the flux of cosmic gamma rays at the top of the atmosphere,
and $\tau_{tot}$ is the cross section of the total attenuation per unit mass.
On the other hand,
$f_s (z, \theta)$ is the flux of cosmic gamma rays
scattered into the aperture of the detector 
after one or more Compton scatterings in the overlying atmosphere.

\begin{figure}
	\plotone{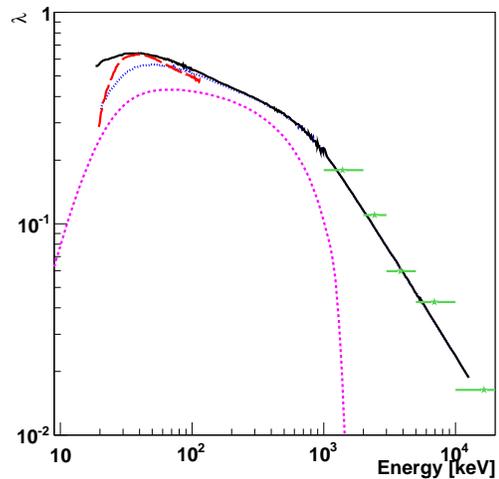}
	\caption{
		Ratio of the scattered component $\lambda$ of diffuse cosmic gamma rays to the total (scattered plus nonscattered) gamma rays 
		as a function of the energy 
		at zenith angles between $0$ and $20$ degrees
		at an atmospheric depth of $8.0$ g cm$^{-2}$.
		The solid and dotted lines show our simulation results 
		with and without Rayleigh scattering, respectively.
		The long-dashed line, dashed line, and stars show 
		the simulation by \citet{Horstman_1971}, 
		the calculation by \citet{Makino_1970}, 
		and the simulation by \citet{Schonfelder_1977}, respectively.
		Each model is at an atmospheric depth of $8.0$ g cm$^{-2}$.
		\label{fig:correction_factor}
	}
\end{figure}

\begin{figure}
	\plotone{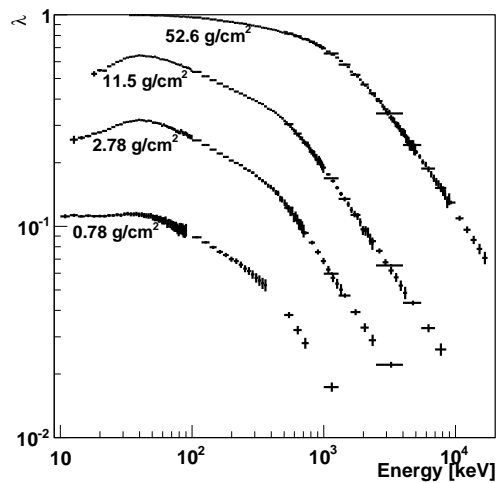}
	\caption{
		Our simulation results with Rayleigh scattering of the ratio of the scattered component 
		$\lambda$ of diffuse cosmic gamma rays as a function of the energy 
		at zenith angles between $0$ and $20$ degrees
		at atmospheric depths of $52.6$, $11.5$, $2.78$, and $0.78$ g cm$^{-2}$ (upper to lower lines).
		There are five data sets for each curve: $10$--$100$ keV, $100$ keV--$1$ MeV, $500$ keV--$5$ MeV, $1$--$10$ MeV, and $2.5$--$25$ MeV,
		and the points in overlapped energy range  are overlaid.
		\label{fig:c_factor_various_depth}
	}
\end{figure}

\begin{figure}
	\plotone{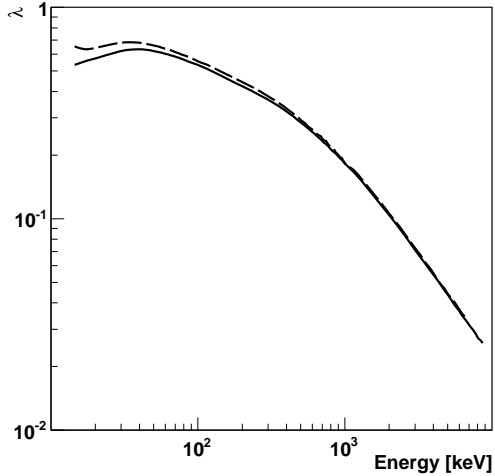}
	\caption{
		Scattering component ratio $\lambda$
		for the initial spectra with power indices of $2.0$ (solid line) and $2.5$ (dashed line)
		at an atmospheric depth of $7.0$ g cm$^{-2}$ and zenith angles between $0$ and $60$ degrees.
		\label{fig:c_factor_index}
	}
\end{figure}

\begin{figure}
	\plotone{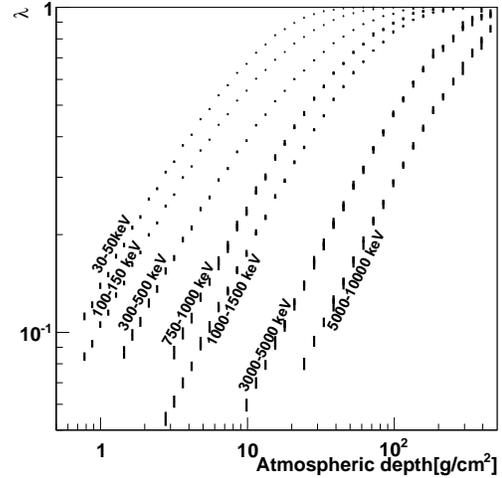}
	\caption{
		Ratio of the scattering component $\lambda$ as a function of the atmospheric depth
		at zenith angles between $0$ and $5$ degrees,
		in the energy ranges $30$--$50$ keV, $100$--$150$ keV, $300$--$500$ keV, $750$--$1000$ keV,
		$1000$--$1500$ keV, $3000$--$5000$ keV, and $5000$--$10000$ keV.
		\label{fig:c_factor_depth}
	}
\end{figure}

\begin{figure}
	\plotone{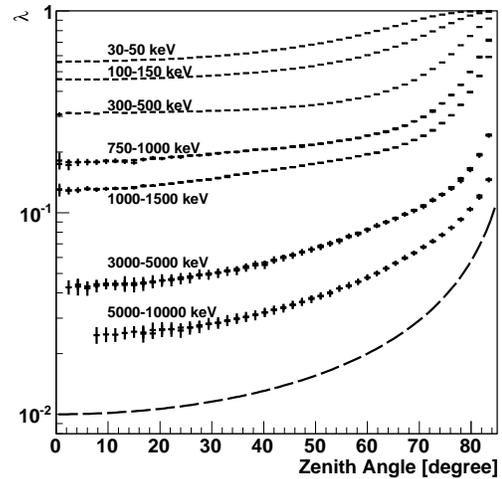}
	\caption{
		Ratio of the scattering component $\lambda$ as a function of the zenith angle
		in the energy ranges of $30$--$50$ keV, $100$--$150$ keV, $300$--$500$ keV,
		$750$--$1000$ keV, $1000$--$1500$ keV, $3000$--$5000$ keV, and $5000$--$10000$ keV.
		The dashed line is proportional to $1/\cos \theta$. 
		\label{fig:c_factor_zenith}
	}
\end{figure}

To obtain the ratio of the scattered component to the total diffuse cosmic gamma-ray flux, 
we simulated the transport of gamma rays in the atmosphere with Geant4. 
We defined $80$ atmospheric layers from sea level to an altitude of $80$ km, 
where each layer was a $1$-km thick sphere, 
filled with air with the density appropriate for each altitude. 
Initial gamma rays were generated at random positions on the surface of a sphere of $650$ km radius and given random initial directions, 
and we traced their histories.
Then, the ratio of the scattered component was calculated 
as a function of energy, atmospheric depth, and zenith angle.
Figures~\ref{fig:correction_factor} and \ref{fig:c_factor_various_depth} show 
the scattered component ratio $\lambda (z, \theta)$ as a function of the detected energy,
\begin{equation}
	\lambda (z, \theta)  = \frac{f_s (z, \theta)}{f_n (z, \theta) + f_s (z, \theta)} ,
	\label{eq:correction}
\end{equation}
assuming a photon index of $2.0$ in the primary spectrum.
There are two breaks in Fig.~\ref{fig:correction_factor}. 
The first break is at around $30$ keV, where absorption and scattering dominate below and above the peak energy, respectively.
The energy of the first break is constant at any atmospheric depth, as shown in Fig.~\ref{fig:c_factor_various_depth}. 
The second break in Fig.~\ref{fig:correction_factor} appears at approximately $1$ MeV. 
The energy of the second break decreases as the atmospheric depth decreases, as shown in Fig.~\ref{fig:c_factor_various_depth}, 
and the optical depth at the second break is approximately one. 
Therefore, for gamma rays with energies above the second break, the probability of scattering is so low that $\lambda$ decreases.
Moreover, Rayleigh scattering contributes a large amount (more than $10$\%) to the scattering component ratio under $50$ keV.
Compared with the previous calculation and simulations of the scattered component ratio function, 
our simulation result has a difference of $\leq 10$\% 
from the simulation of Horstman at $30$--$100$ keV \citep{Horstman_1971}, 
$\sim 20$\% from the calculation of Makino at hundreds of keV \citep{Makino_1970}, 
and $\sim 10$\% from the simulation of Sch\"onfelder above $1$ MeV \citep{Schonfelder_1977}.
Makino's calculation assumed 
that the scattering angle dependence of the cross section was negligible, 
and only one scattering occurred. 
Therefore, 
Makino's calculation differs considerably from our result 
in the energy range under $100$ keV because of the scattering number limitation,
and it also differs above $100$ keV 
because it neglects the scattering angle dependence of the cross section.
The contribution of cosmic gamma rays that are scattered in the atmosphere is not negligible in the sub-MeV/MeV region;
thus, we also used Geant4 to simulate the scattering component ratio as a function of energy (for various power indices),
atmospheric depth, and zenith angle (Fig.~\ref{fig:c_factor_index}--~\ref{fig:c_factor_zenith}).
Figure~\ref{fig:c_factor_index} shows that the ratios for different power indices differ by less than $5$\% 
in the range between tens of keV and a few MeV;
thus, the choice of the power index does not strongly affect our results.
Figure~\ref{fig:c_factor_depth} shows that the atmospheric depth dependence of $\lambda $ is similar to that of Makino's calculation, 
$\case{1}{1-\lambda} \propto \log (1 + \tau_{tot} z)$,
although there is a $10$\%--$20$\% difference in the energy range of $100$--$500$ keV.
Figure~\ref{fig:c_factor_zenith} shows that although the zenith dependence of $\lambda$ below a few MeV is weak 
in the region of zenith angles less than $50$ degrees, it changes abruptly above $70$ degrees.
Thus, the zenith angle dependence below a few MeV is not consistent with $1/\cos \theta$, 
where $\theta$ is the zenith angle of the incident gamma ray.
For the higher energies, $\lambda$ is similar to $1/\cos \theta$.

\subsection{Diffuse Cosmic and Atmospheric Gamma-ray Flux}
To study the origin of the reconstructed events, 
we divided the obtained data at atmospheric depths between $5.2$ g cm$^{-2}$ and $168$ g cm$^{-2}$ into $10$ segments as Table~\ref{tab:time_segment}.
Then,
we obtained the gamma-ray flux from the zenith direction ($\theta = 0$--$10$ degrees) for each time segment
by analyzing the reconstructed events with a simulated response matrix 
for incident energies of 100, 150, 200, 300, 400, 511, 600, 1000, and 1500 keV
and incident zenith angles of 0, 20, 40, 60, 80, 120, 150, and 180 degrees. 
In the calculation,
the zenith angle dependences of both diffuse cosmic and atmospheric gamma-ray fluxes must be considered,
because our ETCC had a wide FOV of $3$ sr.
For diffuse cosmic gamma rays, we simulated the zenith angle dependence as described in section~4.3.
For the zenith angle distribution of combined cosmic and atmospheric gamma rays, 
only three models exist and the conditions under which they apply are limited, as follows.
1) In Ling's model, the source function, which is the production rate of atmospheric gamma rays at the generated position, 
is defined on the basis of observed data but has no dependence on the zenith angle.
In transport, only absorption is included, and scattering is not considered.
Ling's model also describes the flux of diffuse cosmic gamma rays between $300$ keV and $10$ MeV at atmospheric depths smaller than $500$ g cm$^{-2}$,
which is comparable to the results of several experiments, 
although the upward gamma-ray flux differs by a factor of three from the observed data \citep{Schonfelder_1977}.
2) Graser's model calculates atmospheric gamma rays based on bremsstrahlung using some models of the electron flux. 
This model ignores scattering in transport and $\pi^0$-decay gamma rays. 
Thus, this model is effective in the energy range of $1$--$30$ MeV.
3) Costa's model is an empirical model for the sum of cosmic and atmospheric gamma rays 
based on observations by \citet{Schonfelder_1977, Schonfelder_1980} and \citet{Ryan_1977}. 
This model fits the experimental data well but is applicable only at atmospheric depths under $10$ g cm$^{-2}$.
Given the limitations of each of these models,
it is difficult to choose which one best describes the zenith angle distribution of combined cosmic and atmospheric gamma rays.

As described in section~4.3,
the contribution of the scattering of cosmic gamma rays in the atmosphere is not negligible within the energy range of SMILE-I.
Similarly, the scattering of atmospheric gamma rays in the atmosphere is not negligible.
However, Graser's model ignores scattering in the atmosphere.
Ling's model also does not include any scattering in transport. 
However, the difference from the observed zenith-angle dependence is less than that of Graser's model, 
because the source function of Ling's model was fitted to the sum fluxes of cosmic and atmospheric gamma rays obtained from previous observations.
Moreover, Ling's model covers the range of atmospheric depths corresponding to all altitudes of the SMILE-I flight.
We, therefore, adopted Ling's model including the cosmic component.
However, the Ling's model describes the flux only above $300$ keV.
Thus, we extended Ling's model, 
which was extrapolated to less than $300$ keV with a single power-law spectrum fitted between $300$ keV and $5$ MeV.
Below an atmospheric depth of $10$ g cm$^{-2}$, 
the difference between the extended Ling's model and Costa's model in the energy range below $300$ keV is less than $\pm 5$\% at all zenith angles.
However, we could not find a report on any observational data below $300$ keV 
that could be compared with the extended Ling's model above the atmospheric depth of $10$ g cm$^{-2}$.
Therefore, although we adopted an extended Ling's model,
we could not estimate its uncertainty.

\begin{figure}
	\plotone{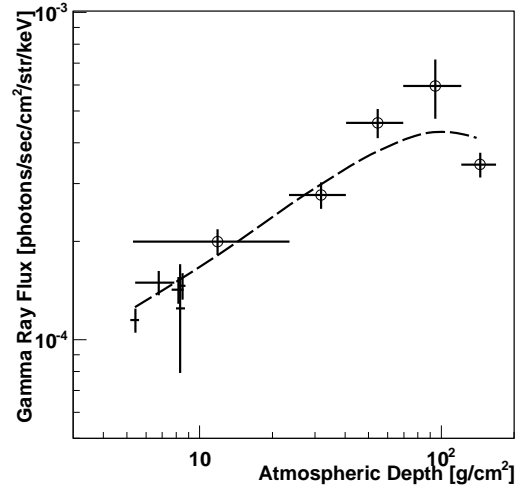}
	\caption{
		Reconstructed gamma-ray flux in the zenith angle range of $0$ to $10$ degrees
		in the energy range $125$--$1250$ keV as a function of atmospheric depth. 
		The circles represent data taken with the balloon ascending.
		The dashed line shows the sum of 1) the simulated background (solid line in Fig.~\ref{fig:bg_depth}),
		2) the diffuse cosmic gamma-ray flux of our extended Ling's model between $125$ and $1250$ keV,
		and 3) the atmospheric gamma-ray flux of the same model.
		The atmospheric component of the extended Ling's model is scaled by the cutoff rigidity.
		\label{fig:growth_curve}
	}
\end{figure}

\begin{deluxetable*}{cccccc}
	\tablecolumns{6}
	\tablewidth{0pc}
	\tablecaption{Obtained gamma-ray fluxes in the zenith angle range of $0$--$10$ degrees for each time segment. \label{tab:time_segment}}
	\tablehead{
		\colhead{Time} &\colhead{Atmospheric depth} &\multicolumn{4}{c}{Flux [$10^{-4}$ photons s$^{-1}$ cm$^{-2}$ sr$^{-1}$ keV$^{-1}$]} \\
		\colhead{[JST]} &\colhead{[g cm$^{-2}$]} &\colhead{125--250 keV} &\colhead{250--550 keV} &\colhead{550--1250 keV} &\colhead{125--1250 keV}
	}
	\startdata
	07:20--07:47 &145 (121--168)	&$8.8\pm1.2$ &$5.5\pm0.6$ &$1.6\pm0.2$	&$3.4\pm0.3$\\
	07:47--08:00 &95	(70--121)	&$18\pm10$   &$9.0\pm1.2$ &$2.6\pm0.4$	&$6.0\pm1.2$\\
	08:00--08:12 &55 (40--70)		&$12\pm 2$   &$7.4\pm1.0$ &$2.1\pm0.4$	&$4.6\pm0.5$\\
	08:12--08:24 &32 (23--40)		&$7.1\pm1.1$ &$4.5\pm0.5$ &$1.3\pm0.2$	&$2.8\pm0.3$\\
	08:24--08:56 &12 (5.3--24)		&$5.1\pm0.7$ &$3.2\pm0.4$ &$0.92\pm0.13$	&$2.0\pm0.2$\\
	08:56--10:08 &5.4	(5.2--5.6)		&$3.0\pm0.4$ &$1.9\pm0.2$ &$0.53\pm0.07$	&$1.1\pm0.1$\\
	10:08--11:00 &6.8	(5.4--7.9)		&$3.9\pm0.5$ &$2.4\pm0.3$ &$0.69\pm0.09$	&$1.5\pm0.1$\\
	11:00--11:30 &8.2	(7.7--8.6)		&$3.7\pm0.5$ &$2.3\pm0.3$ &$0.65\pm0.10$	&$1.4\pm0.1$\\
	11:30--12:06 &8.5	(8.3--8.7)		&$3.8\pm0.6$ &$2.4\pm0.3$ &$0.67\pm0.10$	&$1.5\pm0.1$\\
	12:33--12:59 &8.3	(8.0--8.7)		&$3.2\pm2.3$ &$2.0\pm1.2$ &$0.57\pm0.33$	&$1.2\pm0.5$
	\enddata
\end{deluxetable*}
We used this extended Ling's model of the cosmic and atmospheric gamma rays along with the simulated response matrices 
to unfold the observed events with reconstructed directions  in the zenith angle below $60$ degrees,
resulting in the gamma-ray fluxes listed in Table~\ref{tab:time_segment}.  
In this analysis, we assumed  
that the zenith angle dependence of the sum of diffuse cosmic, atmospheric, and  instrumental gamma rays was the same as the extended Ling's model.
Using this data, 
the growth curve, which is the dependence of the gamma-ray flux on atmospheric depth, in the energy range of $125$--$1250$ keV was obtained,
as shown in Fig.~\ref{fig:growth_curve}.
For comparison with the extended Ling's model, 
the atmospheric gamma-ray flux was considered to have a dependence on $R_{cut}^{-1.13}$, 
where $R_{cut}$ is the cutoff rigidity \citep{Thompson_1981}. 
Because the cutoff rigidities of the SMILE-I flight and Ling's model are $9.7$ GV and $4.5$ GV, respectively, 
the sum of the atmospheric gamma-ray flux of the extended Ling's model multiplied by $(9.7/4.5)^{-1.13}$
and the diffuse cosmic gamma-ray flux of the extended Ling's model between $125$ keV and $1250$ keV
are also shown in Fig.~\ref{fig:growth_curve}. 
The SMILE-I result was consistent with the sum of the extended Ling's atmospheric and cosmic model 
and the background gamma-ray flux represented by the solid line in Fig.~\ref{fig:bg_depth}.
At atmospheric depths from $5$ to $100$ g cm$^{-2}$, 
the variability of the ratio of the background particle flux to the cosmic and atmospheric gamma-ray flux is slight ($< \pm 5$\%).
Thus, the ratio of the non-Compton-scattering events of neutrons or charged particles to the reconstructed events is 
approximately equal to that of the level flight, $2$\%, which was estimated in section 4.2.
In other words, $98$\% of the reconstructed events were induced by cosmic, atmospheric, and instrumental background gamma rays.
Therefore, we calculated the fluxes of cosmic and atmospheric gamma rays by using the fluxes obtained from the SMILE-I observations 
assuming that the reconstruction of non Compton-scattering events was negligible.

\begin{figure}
	\plotone{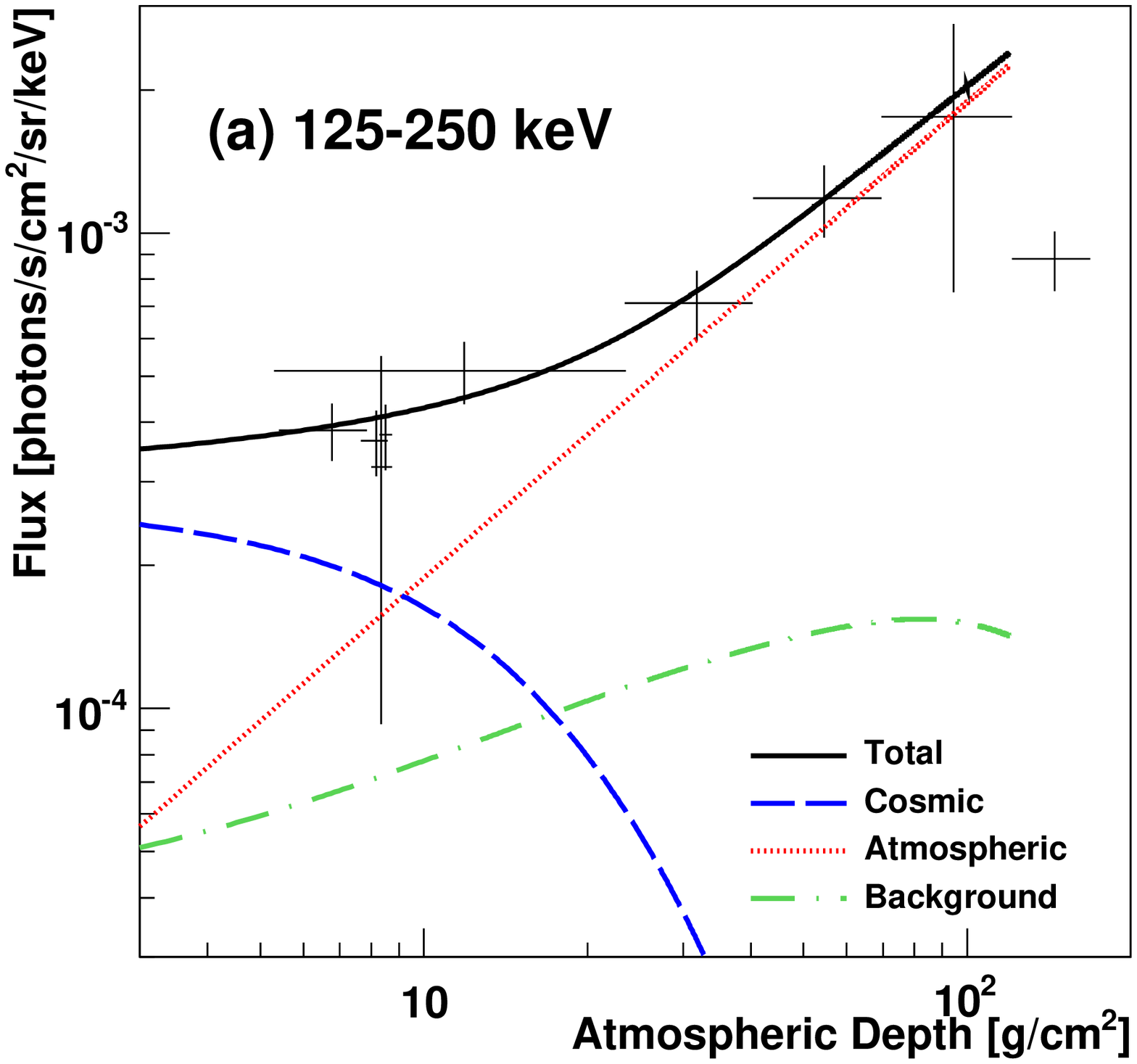}
	\plotone{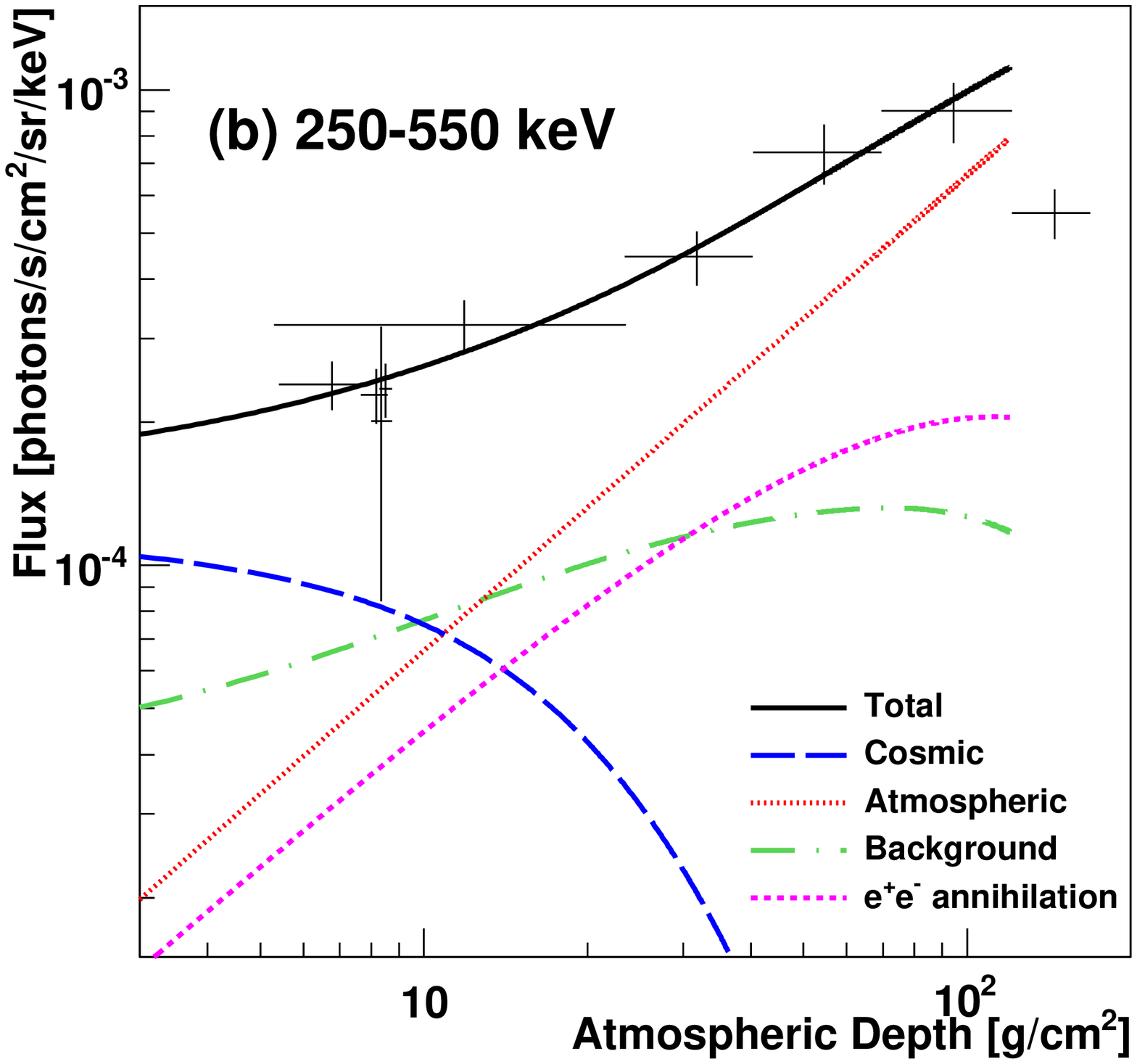}
	\plotone{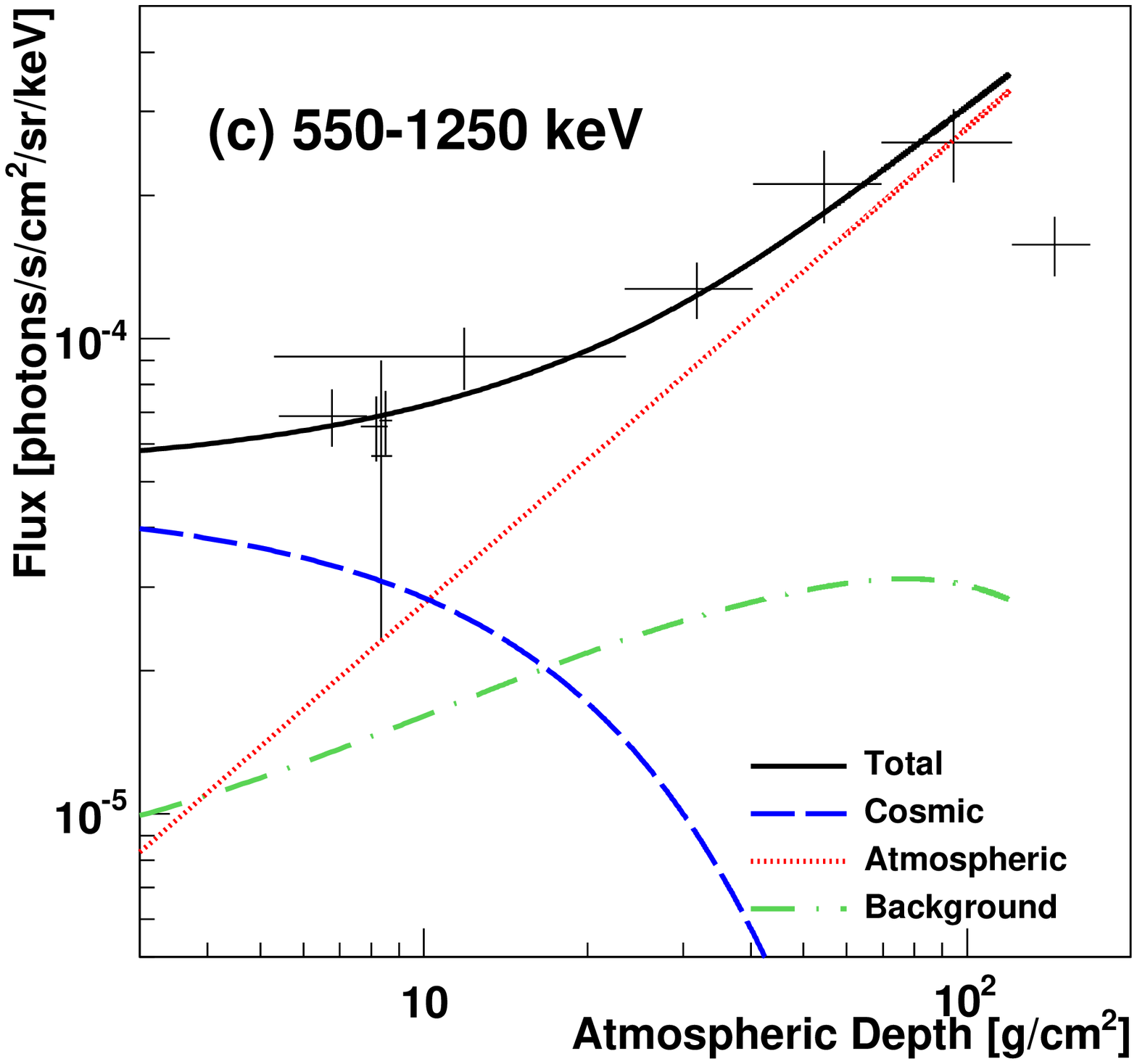}
	\caption{
		Separation of atmospheric and cosmic components of gamma rays 
		by fitting the growth curves in the energy ranges (a) $125$--$250$ keV,
	    (b) $250$--$550$ keV, and (c) $550$--$1250$ keV,
	    in the zenith angle range of $0$--$10$ degrees.
		The long-dashed, dotted, dot-dashed, and dashed lines represent 
		the contributions from cosmic gamma rays, atmospheric gamma rays, instrumental background components,
		and the electron--positron annihilation line, respectively.
		The solid line represents their sum.
		\label{fig:growth_fit}
	}
\end{figure}
\begin{figure*}
	\plotone{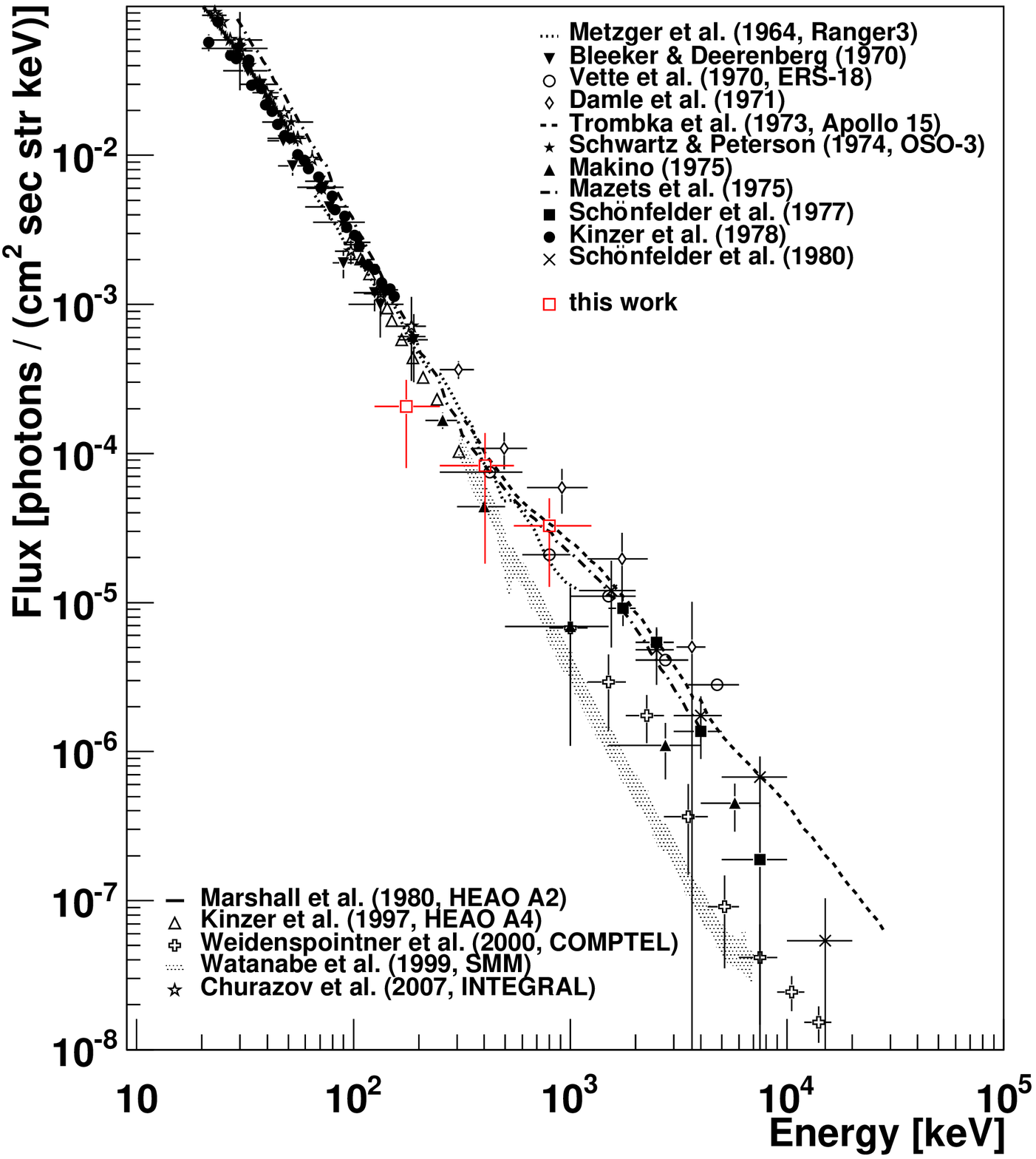}
	\caption{
		Differential fluxes of diffuse cosmic gamma rays as a function of energy.
		Open squares denote our results, and the uncertainty bars represent the $1\sigma$ combined statistical and systematic uncertainties.
		References to previous observations are provided in the text.
		\label{fig:cosmic_flux}
	}
\end{figure*}
\begin{figure*}
	\plotone{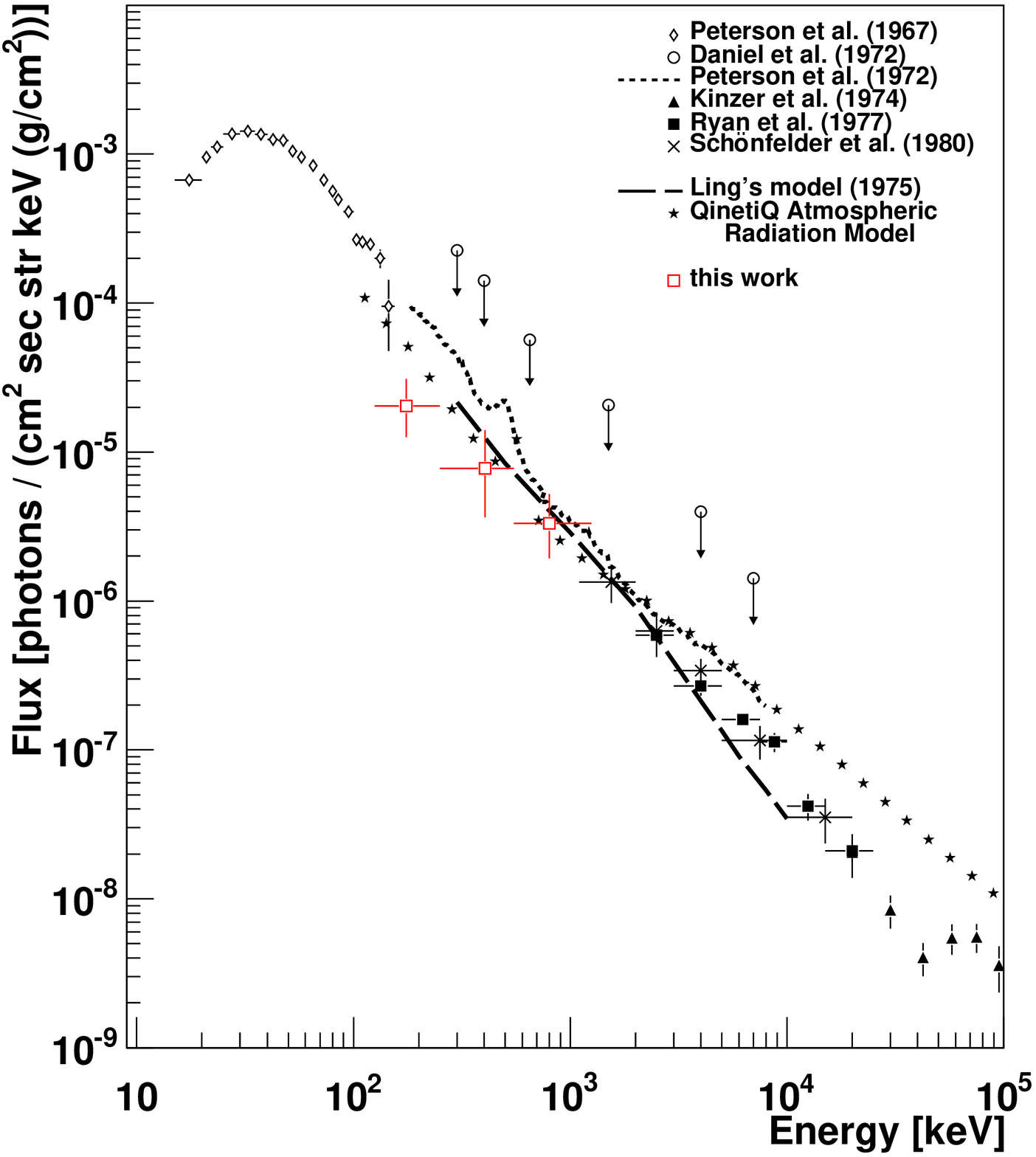}
	\caption{
		Differential fluxes of downward atmospheric gamma rays in the zenith direction ($0$--$10$ degrees), multiplied by $(R_{cut}/9.7 \ \rm{GV})^{1.13}$
		as a function of energy.
		Open squares denote our results,
		and the uncertainty bars represent the $1\sigma$ combined statistical and systematic uncertainties.
		The dashed line and stars represent Ling's model and QARM, respectively.
		References to previous observations are provided in the text.
		\label{fig:atmos_flux}
	}
\end{figure*}

\begin{deluxetable*}{ccc}
	\tablecolumns{3}
	\tablewidth{0pc}
	\tablecaption{
		Obtained diffuse cosmic and atmospheric gamma-ray fluxes in the zenith direction ($0$--$10$ degrees). 
		\label{tab:flux_eng}
	}
	\tablehead{
		\colhead{Energy} &\colhead{Cosmic gamma rays\tablenotemark{$\dagger$}} &\colhead{Atmospheric gamma rays\tablenotemark{$\dagger$}} \\
		\colhead{[keV]} &\colhead{[ph s$^{-1}$ cm$^{-2}$ sr$^{-1}$ keV$^{-1}$]} 
			&\colhead{[ph s$^{-1}$ cm$^{-2}$ sr$^{-1}$ keV$^{-1}$ (g cm$^{-2}$)$^{-1}$]}
	}
	\startdata
	125--250	&$2.1 \pm 0.7 ^{+ 1.0}_{- 0.8} \times 10^{-4}$	&$2.0 \pm 0.6 ^{+ 0.5}_{- 0.9} \times 10^{-5}$\\
	250--550	&$8.3 \pm 3.4 ^{+ 5.5}_{- 4.3} \times 10^{-5}$	&$7.8 \pm 2.9 ^{+ 3.0}_{- 5.5} \times 10^{-6}$\\
	550--1250	&$3.3 \pm 1.2 ^{+ 1.6}_{- 1.2} \times 10^{-5}$	&$3.3 \pm 1.0 ^{+ 0.9}_{- 1.6} \times 10^{-6}$
	\enddata
	\tablenotetext{$\dagger$}{The first uncertainty term is statistical and the second is systematic.}
\end{deluxetable*}
To obtain the fluxes of diffuse cosmic gamma rays and atmospheric gamma rays, 
we used the difference in the atmospheric depth dependence of each component.
For the cosmic gamma-ray flux at an atmospheric depth of $z$, 
we assumed 
\begin{equation}
	f_c (z) = \frac{\eta}{1 - \lambda (z, 0)} \exp (- \tau_{tot} z) ,
\end{equation}
with equations (\ref{eq:cosmic}), (\ref{eq:cs_att}), and (\ref{eq:correction}) at $\theta = 0$.
On the other hand,
the atmospheric gamma-ray flux $f_a (z)$ is nearly proportional to the atmospheric depth \citep{Schonfelder_1977},
\begin{equation}
	f_a (z) = \xi z .
\end{equation}
Then, we fitted the observed growth curve to
\begin{equation}
	f(z) = \xi^\prime z + \frac{\eta^\prime}{1 - \lambda (z, 0)} \exp ( -\tau_{tot} z) + f_b (z) + F_{511} (z) \delta (E_{0} - 511 \ \rm{[keV]}) ,
	\label{eq:growth}
\end{equation}
where $f_b$ is the flux of background photons estimated by the Geant4 simulation (Fig.~\ref{fig:bg_depth}),
$F_{511}$ is the flux of the atmospheric annihilation line described by equation~(1) in \citet{Ling_1977},
and $\xi^\prime$ and $\eta^\prime$ are the attenuated fluxes through the pressurized vessel.
The attenuated fluxes are given as $\xi^\prime = \xi \exp ( -\tau_{abs} d l)$ and $\eta^\prime = \eta \exp ( -\tau_{abs} d l)$,
where $\tau_{abs}$ is the cross section of photoelectric absorption in aluminum,
$d = 2.7$ g cm$^{-3}$ and $l = 3$ mm are the density and the thickness of the pressurized vessel, respectively.
The growth curves with the fitting functions are shown in Fig.~\ref{fig:growth_fit},
and these functions serve to isolate individual contributions to the total flux 
so that separate fluxes for diffuse cosmic gamma rays and atmospheric gamma rays can be obtained.
However, at larger atmospheric depths than approximately $100$ g cm$^{-2}$, 
we have a problem because the gamma-ray flux is not proportional to the atmospheric depth, as shown by Ling's model in Fig.~\ref{fig:growth_curve}.
Therefore, the data at the highest atmospheric depth was not used in the fitting. 
The obtained fluxes of cosmic and atmospheric gamma rays are listed in Table~\ref{tab:flux_eng}.
The systematic uncertainties in the cosmic and atmospheric gamma-ray fluxes are caused by the following:
1) The uncertainty of $15$\% in the detection efficiency (section 2).
2) The difference of $5$\% in the relative ratio of the zenith angle distributions between the extended Ling's model and Costa's model (section 4.4).
3) The uncertainty of $20$\% in the background gamma-ray flux obtained by varying the parameters in QARM (section 4.2).
4) The uncertainty of $5$\% in the correction factor of scattering in the atmosphere (section 4.3).
Figures~\ref{fig:cosmic_flux} and \ref{fig:atmos_flux} show 
the differential flux of diffuse cosmic gamma rays and atmospheric gamma rays, respectively,
with results of previous observations for comparison. 
In Fig.~\ref{fig:atmos_flux}, fluxes from observations for which the FOV was not described in the references were not plotted.
In the energy range of $125$--$250$ keV, the flux of atmospheric gamma rays obtained by SMILE-I 
had a difference of $3$ sigma from the observation of \citet{Peterson_1972}. 
However, other results for the diffuse cosmic and atmospheric fluxes obtained by SMILE-I were consistent 
with previous observations and the Ling and QARM models of atmospheric gamma rays within $2$ sigma.

\subsection{Sensitivity and Future Prospects}
\begin{figure}
	\plotone{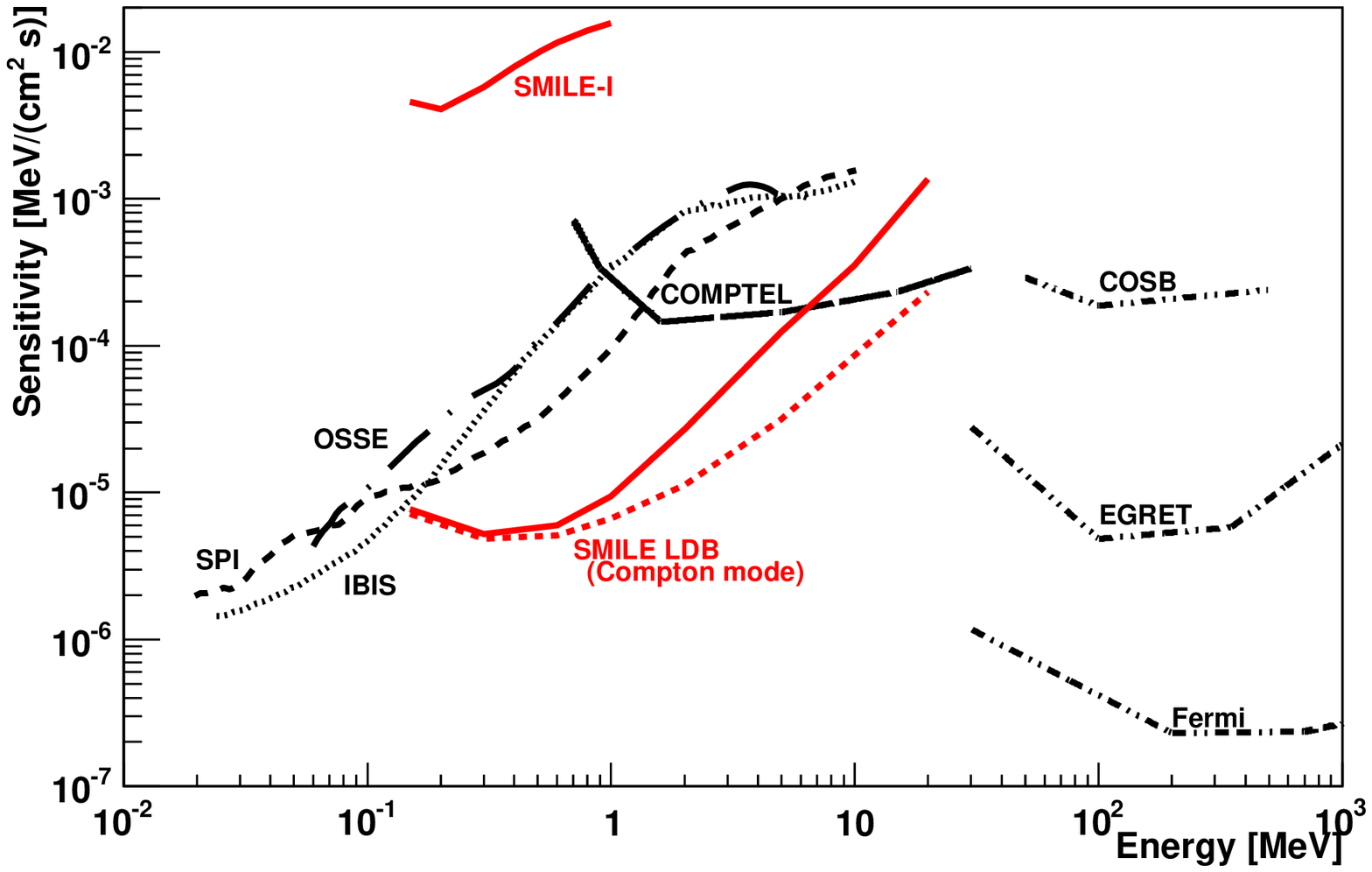}
	\caption{
		Continuum detection sensitivity as a function of energy with an observation time of $10^6$ s at a significance of $3 \sigma$. 
		For comparison with previous observations, the sensitivities of SMILE were overlaid on Fig.~1 of \citet{Schonfelder_2004}.
		The solid and short-dashed lines represent 
		the sensitivity of an ETCC with a volume of $50 \times 50 \times 50$ cm$^3$ aboard a long-duration balloon (LDB) through Compton scattering,
		without and with an absorber of Compton-recoil electrons, respectively.
		\label{fig:sensitivity}
	}
\end{figure}

Using the SMILE-I result, we estimated the continuum detection sensitivity of this camera. 
Diffuse cosmic and atmospheric gamma rays, and those induced by charged particles and neutrons,
which were observed in the SMILE-I experiment, would form the background when we observe a celestial point source.
The minimal detectable flux $F_{min}$ at the significance of $3 \sigma$ is described as
\begin{equation}
	F_{min} = 3 \ \sqrt{\frac{f \ \Delta E \ \Delta \Omega}{A_{eff} \ T_{obs}}} ,
\end{equation}
where $f$ is the energy spectrum of background radiation in units of photons s$^{-1}$ cm$^{-2}$ sr$^{-1}$ MeV$^{-1}$, 
and $A_{eff}$, $T_{obs}$, $\Delta E$, and $\Delta\Omega$ are 
the effective area, effective observation time, energy resolution, and angular resolution, respectively \citep{Schonfelder_2001}.
From the results of the SMILE-I flight,
the sensitivity of SMILE-I is expected to reach $4.1 \times 10^{-3}$ MeV s$^{-1}$ cm$^{-2}$ at $300$ keV
and $1.2 \times 10^{-2}$ MeV s$^{-1}$ cm$^{-2}$ at $600$ keV, assuming $\Delta E = E$ and $T_{obs} = 10^6$ s (Fig.~\ref{fig:sensitivity}).

For the next flight (hereafter SMILE-II) to observe the Crab Nebula from balloon altitudes, 
the ETCC needs upgrading in both the angular resolution and effective area. 
As described in equation~(\ref{eq:phi}), 
the ARM, which is the accuracy of the scattering angle, depends mostly on the energy resolution of the absorber. 
Therefore, improving the energy resolution of the scintillation camera 
would give the ETCC better angular resolution.
A good ARM resolution of $4$ degrees (FWHM) at $662$ keV was realized 
using a recently developed scintillator, LaBr$_3$:Ce, 
with an excellent energy resolution of $3.0$\% and $5.8$\% at $662$ keV 
for a monolithic crystal and an array of pixels, respectively \citep{Kurosawa_2009, Kurosawa_2010}.
From this result, the expected ARM resolution would be $1.7$ degrees and $1.3$ degrees (FWHM) at $2$ MeV and above $5$ MeV, respectively.
As for the effective area,
we have developed a larger ETCC based on a TPC with a volume of $30 \times 30 \times 30$ cm$^3$ \citep{Miuchi_2007}, 
and its gamma-ray imaging has been successful \citep{Ueno_2008}.
With these improvements and the assumption of the instrumental background scaled by the geometrical area of SMILE-I, 
it is expected that the ETCC for SMILE-II on a mission to detect the Crab Nebula during a level flight of $10^4$ s
will have approximately ten times higher sensitivity than that of SMILE-I. 

We have additional ideas for improving the future ETCC.
First, we are considering inserting an electron absorber between the electron tracker and the scintillators.
In the MeV region, 
because the energy of Compton-recoil electrons is as high as an MeV, 
the Compton-recoil electrons partly escape from the effective volume of the TPC. 
If an absorber that measures the energy of the escaping Compton-recoil electrons is placed between the TPC and scintillators,
the sensitivity in the MeV region can be improved.
Inserting the electron absorber has the added benefit of reducing the coincidence-timing window, 
because we can take the coincidence between the electron absorber and scintillators. 
Second, we will adopt CF$_4$ gas as the electron tracker. 
Compared with Xe gas, the electron diffusion during electron drift and the multiple scattering effect of the recoil electron are both smaller. 
Thus, we will be able to trace the electron tracks more accurately. 
In addition, the position resolution of the Compton points and the accuracy of the recoil direction should be improved by updating the tracking algorithm.
Last, we will develop the reconstruction algorithm on the basis of a reflexive calculation, such as the likelihood or maximum entropy method.
If we realize an ETCC with a high ARM due to the use of LaBr$_3$ PSAs with a pixel size of $6 \times 6 \times 40$ mm$^3$,
equipped with a TPC with a larger volume of $50 \times 50 \times 50$ cm$^3$ filled with CF$_4$ gas at a higher pressure of $2$ atm, 
the sensitivity will become ten times better than that of COMPTEL in the sub-MeV region, as shown in Fig.~\ref{fig:sensitivity}.
Moreover, 
when a photon with energy above $1.02$ MeV causes electron--positron pair creation in the TPC, 
we can obtain the incident photon's momentum by measuring the momenta of both the electron and positron with the TPC \citep{Orito_2003}. 
Therefore, the camera will have better sensitivity above a few MeV than that shown in Fig.~\ref{fig:sensitivity}.
In the near future, 
we will experiment on gamma-ray reconstruction by electron--positron pair creation by using an ETCC and verify the result with a simulation,
and also estimate the sensitivity of an ETCC onboard a satellite
where the radiation environment is much different from that at balloon altitudes, 
by a performing the full-scale Monte Carlo simulation.

\section{Conclusion}
We are developing an ETCC
as an MeV gamma-ray telescope to conduct an all-sky survey in the next generation 
with sensitivity that is one order of magnitude higher than that of COMPTEL.
To prepare for future observations onboard a satellite,
we performed a balloon experiment, SMILE.
As the first step of SMILE (SMILE-I),
we launched an ETCC
consisting of a gaseous TPC with a volume of $10 \times 10 \times 14$ cm$^3$ and GSO:Ce scintillation cameras onboard a balloon in 2006 
for the observation of diffuse cosmic gamma rays and atmospheric gamma rays.
In this flight, 
we successfully detected $2.1 \times 10^3$ reconstructed events from all directions,
$420$ of which were detected in an FOV of $3$ sr during the live time of $3.0$ hr of level flight. 
To evaluate the instrumental background,
we estimated particles incident on the detector with simulations using QARM and Geant4; 
the reconstructed events included $1.7$\% neutron events, $0.23$\% charged-particle events, and $98$\% gamma-ray events.
Because the contribution of cosmic gamma rays that are scattered in the atmosphere is not negligible in the sub-MeV/MeV region,
we also use Geant4 to simulate the transport in the atmosphere as a function of energy, zenith angle, and atmospheric depth.
Using these simulations, 
we analyzed the growth curve in the $125$--$1250$ keV range 
and obtained the fluxes of diffuse cosmic gamma rays and atmospheric gamma rays. 
Our results are consistent with those of the previous experiments.
The flight of SMILE-I demonstrated 
that our ETCC can perform gamma-ray selection and powerful background rejection,
and thus, may improve MeV gamma-ray astronomy by having better continuum sensitivity.
As the next step, 
we plan to observe the Crab Nebula by using an ETCC with a $30 \times 30 \times 30$ cm$^3$ TPC to demonstrate gamma-ray imaging. 
Subsequently, we plan to observe several celestial sources with a long-duration balloon flight, 
and finally aim for an all-sky survey at energies between $150$ keV and $20$ MeV 
with sensitivity up to an order of magnitude better than the current and previous missions,
using an ETCC with ARM resolution of $2$ degrees (FWHM) in the MeV region 
and a detection volume of $50 \times 50 \times 50$ cm$^3$ onboard a long-duration balloon or satellite.

\acknowledgments
The authors thank the anonymous referee for carefully reading the manuscript 
and providing helpful comments to improve it.
We also thank the staff of the Scientific Balloon Laboratory, ISAS/JAXA,
for their excellent support during launch, flight, and payload recovery. 
This study was supported by 
a Grant-in-Aid for Scientific Research 
and a Grant-in-Aid from the Global COE program
``Next Generation Physics, Spun from Universality and Emergence''
from the Ministry of Education, Culture, Sports, Science and Technology (MEXT) of Japan,
and the Japan Society for Promotion of Science for Young Scientists.

\end{document}